\documentclass[11pt]{article}
\usepackage{amssymb,fullpage,amsmath,graphicx,latexsym,amsthm,cite,empheq,float,jheppub,url,color}
\usepackage[usenames,dvipsnames]{xcolor}
\usepackage[final]{pdfpages}
\numberwithin{equation}{section}
 \DeclareMathOperator{\kev}{keV} \DeclareMathOperator{\mev}{MeV} \DeclareMathOperator{\gev}{GeV}  \DeclareMathOperator{\cm}{cm}  \DeclareMathOperator{\g}{g} \DeclareMathOperator{\erg}{erg} \DeclareMathOperator{\km}{km}  \DeclareMathOperator{\s}{s}    \DeclareMathOperator{\few}{few} 
 \newcommand{\cC}{{\cal C}}      \newcommand{\cL}{{\cal L}} \newcommand{\cM}{{\cal M}}  \newcommand{\cO}{{\cal O}}    
\newcommand{\ep}{\epsilon}
\newcommand{\epr}{\epsilon_{\rm pr}}
\newcommand{\etr}{\epsilon_{\rm tr}}
\newcommand{\epm}{\epsilon_{\rm m}}
 
  \newcommand{\eg}{{\it e.g.~}}
\newcommand{\pt}{\partial}   \def\oL{\overline} 
\newcommand{\pL}{\left(} \newcommand{\pR}{\right)} \newcommand{\bL}{\left[} \newcommand{\bR}{\right]} \newcommand{\cbL}{\left\{} \newcommand{\cbR}{\right\}} \newcommand{\mL}{\left|} \newcommand{\mR}{\right|}
\newcommand{\beq}{\begin{equation}} \newcommand{\eeq}{\end{equation}}
\newcommand{\bea}{\begin{eqnarray}} \newcommand{\eea}{\end{eqnarray}}
\newcommand{\alg}[1]{\begin{align} \begin{split} #1 \end{split}  \end{align}}

\newcommand{\tenx}[1]{\times 10^{#1}}
\newcommand{\Eq}[1]{Eq.~(\ref{#1})} \newcommand{\Eqs}[2]{Eqs.~(\ref{#1}) and (\ref{#2})} \newcommand{\Eqm}[2]{Eqs.~(\ref{#1}) through (\ref{#2})}
\newcommand{\Sec}[1]{Sec.~\ref{#1}}  
\newcommand{\Fig}[1]{Fig.~\ref{#1}} 
\newcommand{\Tab}[1]{Tab.~\ref{#1}}
\newcommand{\App}[1]{App.~\ref{#1}}
   \DeclareMathOperator{\re}{Re} \DeclareMathOperator{\im}{Im}

\begin{document}
\title{Revisiting Supernova 1987A Constraints on Dark Photons}
\author{Jae Hyeok Chang, Rouven Essig, and Samuel D. McDermott}
\affiliation{C.~N.~Yang Institute for Theoretical Physics, Stony Brook, NY, USA}\subheader{\rm YITP-SB-16-44}

\abstract{
We revisit constraints on dark photons with masses below $\sim 100$~MeV from the observations of Supernova 1987A. If dark photons are produced in sufficient quantity, they reduce the amount of energy emitted in the form of neutrinos, in conflict with observations.  
For the first time, we include the effects of finite temperature and density on the kinetic-mixing parameter, $\epsilon$, in this environment. 
This causes the constraints on $\epsilon$ to weaken with the dark-photon mass below $\sim 15$~MeV. 
For large-enough values of $\epsilon$, it is well known that dark photons can be reabsorbed within the supernova. 
Since the rates of reabsorption processes decrease as the dark-photon energy increases, we point out that dark photons with energies above the Wien peak can escape without scattering, contributing more to energy loss than is possible assuming a blackbody spectrum.
Furthermore, we estimate the systematic uncertainties on the cooling bounds by deriving constraints assuming one analytic and four different simulated  
temperature and density profiles of the proto-neutron star.    
Finally, we estimate also the systematic uncertainty on the bound by varying the distance across which dark photons must propagate from their point of production to be able to affect the star.  
This work clarifies the bounds from SN1987A on the dark-photon parameter space. 
}
\maketitle

\section{Introduction}\label{sec:intro}
Dark sectors --- containing particles that are not charged directly under the Standard Model forces --- could provide an explanation for various 
shortcomings of the Standard Model, including dark matter candidates, neutrino masses, and the baryon asymmetry.  
Three ``portals'' allow for renormalizable interactions between the Standard Model and a dark sector: the neutrino portal, the Higgs portal, and the vector portal. 
In this paper we focus on the vector portal, in which (at low energies) the Standard Model photon interacts with a new gauge boson 
(the ``dark photon'') through kinetic mixing.  
Dark photons can easily remain hidden from conventional searches despite having masses well below the Weak scale.  
Over the last few years, much progress has been made to map out the viable parameter space and identify promising new search 
directions~\cite{Jaeckel:2010ni,Hewett:2012ns,Essig:NLWCP:2013,Alexander:2016aln}. 

In this paper, we will focus on constraints on dark photons from the observation of a core-collapse supernova in 1987 
in the Large Magellanic Cloud, called SN1987A.  
In the standard picture, the vast majority of energy liberated from the collapsing star, about 99\% of the difference in the gravitational binding energy of the progenitor and the remnant, leaves the supernova in the form of neutrinos~\cite{Raffelt:1996wa}.   
However, if new particles are (a) produced in large numbers and (b) able to travel macroscopic distances without transferring energy back to the stellar material, they provide a cooling mechanism that competes with the neutrinos.  
Since the neutrino burst observed in association with the explosion agrees qualitatively with predictions based on Standard Model-only simulations
~\cite{Burrows:1986me,Burrows:1987zz}, 
such an anomalous energy loss is severely constrained. 
This means that dark-sector particles with masses below the characteristic supernova temperature of $\sim \cO(10$'s of MeV) must be either (a) so weakly coupled they are produced in negligible quantities or (b) coupled strongly enough that their mass and energy gets reprocessed by Standard Model particles. 
The purpose of this paper is to examine this scenario in detail, focusing on a realistic treatment of finite density effects on the Standard Model photon.

We will be interested in a minimal dark sector whose only light particle is a $U(1)'$ gauge boson. This gauge boson can mix with the 
Standard-Model photon, with field strength $F^{\mu\nu}$, in the usual way,
\beq \label{lagrangian}
\cL = \cL_{\rm SM} + \frac12 m'^2 A'_\mu A'^\mu - \frac14F'_{\mu \nu}F'^{\mu \nu} -\frac\ep2 F'_{\mu \nu}F^{\mu \nu}\,,
\eeq
where $F'_{\mu \nu} = \partial_\mu A'_\nu - \partial_\nu A'_\mu $ is the field strength for the dark photon field $A'$ and $\ep$ is the kinetic mixing parameter that allows the dark photon to couple to ordinary matter. The dark photon has a mass $m'$. (We will assign a prime to any quantity associated with the dark photon.)
We assume that the dark photon is the only new particle that could affect the supernova evolution.  
We leave to future work the possibility that there are additional stable, low-mass dark-sector scalars or fermions 
with $U(1)'$ charge~\cite{future-DM-paper} (see also~\cite{Essig:2013vha,izaguirre:2013uxa,Dreiner:2013mua}).  
Such particles, which could constitute the dark matter, could in principle provide another decay channel for dark photons 
and alter the supernova's evolution.
In addition, we also imagine that a dark Higgs boson, which could be responsible for spontaneously breaking the $U(1)'$ gauge symmetry, 
is heavy enough not to effect the supernova evolution. We will eventually set bounds in the $\ep-m'$ parameter space. Early studies indicate $\ep \sim \cO(10^{-10})$ is the order of magnitude of mixing angle that affects the supernova evolution~\cite{Bjorken:2009mm,Dent:2012mx,Kazanas:2014mca,Rrapaj:2015wgs}.

\begin{figure}[t]
\begin{center}
\includegraphics[width=\textwidth]{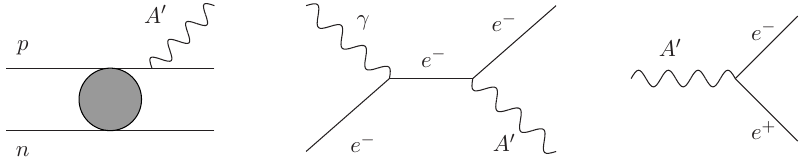}
\caption{Scattering and decay processes involving $A'$ particles: {\bf(left)} bremsstrahlung in neutron-proton ($n-p$) scattering, {\bf(center)} ``semi-Compton'' scattering, and {\bf(right)} decay. The bremsstrahlung diagram is accompanied by two more diagrams at the same order (the $A'$ attached to the incoming $p$ and a charge-exchange interaction). The grey blob in the bremsstrahlung diagram represents non-perturbative $n-p$ scattering. The $A'$ can be produced inside the star by bremsstrahlung and semi-Compton scattering, but inverse versions of bremsstrahlung and semi-Compton contribute to absorption of $A'$ particles on their way out of the star. Decay to $e^+e^-$, possible only for $m'>2m_e$, where $m_e$ is the electron mass, is not in equilibrium due to the low density of positrons in the proto-neutron star.}
\label{feyn-diags}
\end{center}
\end{figure}

Due to the mixing term $\ep F'_{\mu \nu}F^{\mu \nu} /2$ in \Eq{lagrangian}, dark photons can be produced in any process that results in Standard Model photons. The primary processes of interest are depicted as the left two diagrams in \Fig{feyn-diags}.  However, the kinetic mixing angle that dictates the production rate of dark photons can 
be substantially enhanced compared to its value in vacuum due to finite temperature and density effects.  
The Lagrangian parameter and the effective mixing angle are inequivalent when the Standard Model photon receives a nonzero ``plasma mass.'' This is encoded by the polarization tensor $\re \Pi \sim \omega_p^2 =4\pi \alpha_{\rm EM} \sum n_i/E_{F,i}$~\cite{Kapusta:2006pm}, where the sum goes over charge carriers, $n_i$ is their number density, and $E_{F,i}^2 = m_i^2 + (3\pi^2 n_i)^{2/3}$ is their Fermi energy. At high density $n_i \gg m_i^3$, as encountered in the proto-neutron star, contributions to the plasma mass are dominated by the lightest charged particle 
present (i.e.~electrons). The polarization tensor enters the gauge boson mass matrix as an additional mass term, which (dynamically) changes the diagonalization of the kinetic mixing parameter $\ep$. 
The end result is that the mixing angle in medium, $\epm$, depends on $m'$ and $\Pi$. On the mass shell the effective mixing angle is~\cite{An:2013yfc}
\beq \label{in-medium-angle}
\ep^2 \to \epm^2 = \frac{\ep^2}{\pL 1 - \re \Pi/m'^2\pR^2 + \pL\im \Pi/m'^2\pR^2 },
\eeq
where we reserve the symbol $\ep$ to always refer to the mixing angle in vacuum, which is the parameter that enters the Lagrangian. 
The transverse and longitudinal components need to be treated separately: each has its own polarization tensor and hence its own dispersion relation. We discuss how to calculate $\Pi_{L,T}$ in \Sec{sec:A-prime-physics}.

The in-medium mixing angle from \Eq{in-medium-angle} controls both the production and absorption rates of dark photons, so that the amount of energy that flows out of the proto-neutron star is (schematically) $L \sim \omega \Gamma e^{-\tau}$ for a dark photon energy $\omega$, production rate $\Gamma$, and ``optical'' depth $\tau$. As we show below, $\tau$ has a strong dependence on $\omega$, so there are a number of interesting limits for the production rate of the dark photon. We anticipate the results below by noting that at low masses and small mixing the bounds scale such that the product $\ep m'$ is constant, $\ep m'/\mev \lesssim 2\tenx{-9}$, independent of production rate and in contrast to prior results which indicated a flat asymptote of $\ep \lesssim \cO(10^{-10})$ at low mass. 
For large-enough mixing angles the bounds disappear: with $m'< 2m_e$ and $\ep m'/\mev \gtrsim 5\tenx{-5}$, 
the absorption of dark photons through the inverse semi-Compton process is large enough to prevent the anomalous cooling of the supernova, 
while for $m' > 2 m_e$ and $\ep m'/\mev \gtrsim 1\tenx{-6}$, the dark photons rapidly decay to electron-positron pairs and are 
unable to carry away enough energy to affect the supernova evolution.  
We note that the high-$\ep$ limit we obtain in the range $m' < 2m_e$ is up to two orders of magnitude stronger than it would be if we had taken  a blackbody spectrum for escaping particles. Instead, we find that high-energy particles with $\omega \gg \few \times \,T$ are capable of escaping the supernova interior without scattering, leading to relatively higher luminosity compared to that expected from a blackbody spectrum. 

To establish these results, we will need to revisit several different aspects of supernova physics and finite-temperature field theory. In \Sec{SN-details},  we will discuss what is known and what remains uncertain about the explosion of SN1987A. We will focus on the interpretation of the luminosity limits and describe the requirements for transporting energy away from the zone of neutrino production. In \Sec{sec:A-prime-physics}, we will discuss how the $A'$ is produced and propagates in the supernova environment. As is evident from \Eq{in-medium-angle}, a resonance is possible in the mixing angle; we will discuss the circumstances under which this resonance is attained, and what happens when it is suppressed. In \Sec{sec:results}, we will apply these lessons to draw excluded regions in the $m'-\ep$ parameter space, and then we conclude.  Several appendices provide technical details 
of our calculation and a comparison with previous work.

\section{Supernova Constraints on New Particles}
\label{SN-details}

Despite great challenges in modeling the intermediate stages of core-collapse supernovae, the late stages of their evolution are adequately understood at a qualitative level. In particular, observations of the neutrino fluence and energy spectra at Kamioka~\cite{Hirata:1987hu}, the IMB experiment~\cite{Bionta:1987qt}, and Baksan~\cite{Alekseev:1987ej} are in broad agreement with predictions based on simulations of core-collapse supernovae conducted by Lattimer and Burrows~\cite{Burrows:1986me,Burrows:1987zz}. Despite $\cO(1)$ uncertainties on the relevant astrophysics underlying the explosion, discussed at more length below, the agreement between theory and observation illustrates that there is only limited room for the possibility of a competing energy sink during the time of the core collapse. 

This agreement has allowed constraints on a wide variety of models, including axions~\cite{Turner:1987by,Raffelt:1988py}, sterile neutrinos~\cite{Kainulainen:1990bn,Kuflik:2012sw}, and compact extra dimensions~\cite{Hanhart:2000er, Hanhart:2001fx}. Preliminary work on dark photons first appeared in~\cite{Bjorken:2009mm}, followed by calculations of the luminosity in the Born approximation~\cite{Dent:2012mx,Kazanas:2014mca} and then including nonperturbative effects from nucleon-nucleon scattering~\cite{Rrapaj:2015wgs}. In this work, we revisit these constraints, focusing on the effects of finite temperature and density on the limits at low masses. In this section, we focus on how to apply these constraints, and discuss subtleties and uncertainties that hamper a complete analytic understanding of the bounds.

\subsection{Using Luminosity to Obtain Constraints}
\label{lumi}

Dark photons may be emitted as Standard-Model particles collide in the proto-neutron star. If dark photons are readily produced and are able to free-stream out, the total luminosity in dark photons may be so large that it drains the energy that powers electroweak interactions. This could potentially alter the luminosity of neutrinos that provide the terrestrial signal of the supernova explosion. Detailed simulations, including the backreaction of the new heat sink on the Standard-Model bath, are necessary to reliably assess the viability of any given point in the model parameter space. However, a convenient, approximate criterion was established in~\cite{Raffelt:1996wa} by Raffelt: 
if the instantaneous luminosity in novel particles exceeds the value
\beq \label{lum-bound}
L(m',\ep) \geq L_\nu = 3 \times 10^{52} \erg\!/\!\s 
\eeq
when the core reaches peak density $\rho_c \sim 3 \tenx{14}\g/\cm^3$ and temperature $T_c \sim 30\mev$, the duration of the neutrino burst from SN1987A is reduced by half, and the energy spectrum is inconsistent with observations. We will adopt this ``Raffelt criterion'' to find constraints. 
Stronger limits are potentially obtained by bounding the efficiency of the energy transport inside the core or the energy released over the entire neutrino burst, but these would also be subject to much larger systematic uncertainties from not knowing precisely the nature of the progenitor star. 

\begin{figure}[tb]
\begin{center}
\includegraphics[width=.7\textwidth]{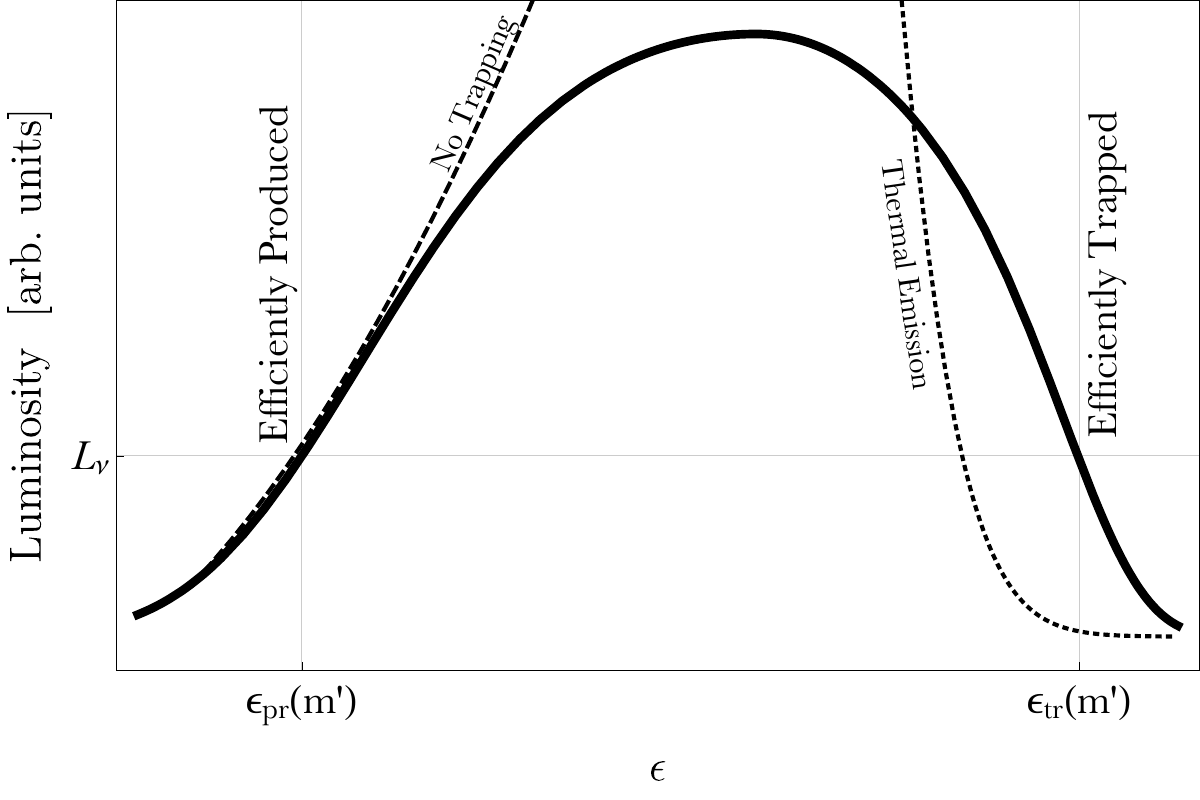}
\caption{Luminosity as a function of mixing angle $\ep$ for a fixed $m'$. Above $\epr(m')$, enough dark photons are produced that their luminosity may exceed $L_\nu$.  For $\ep>\etr(m')$, dark photons produced far in the interior of the proto-neutron star are trapped. 
This means that enough of their energy will be reprocessed into neutrino energy, preventing serious energy depletion. 
The range $\epr(m') \leq \ep \leq \etr(m')$ is ruled out. If $L(m',\ep)$ exceeds $L_\nu$ for no value of $\ep$ at a given mass $m'$, there is no constraint.
If we neglected the optical depth entirely, we would find that the luminosity scales like $\sim \ep^2$,  
shown as a dashed line labelled ``No Trapping''. 
Similarly, if we took the standard assumption that the emission is a blackbody spectrum from a surface at which the average dark photon has unit optical depth, we would find that the luminosity scales like $\sim e^{-\ep^2}$. This underestimates the contribution of very high-energy dark photons and leads to the bound shown with the dotted line labelled ``Thermal Emission''.   
The thermal emission bound is weaker than the real bound.  
}
\label{schematic-luminosity}
\end{center}
\end{figure}

In \Fig{schematic-luminosity} we plot a schematic representation of $L$ for a fixed value of $m'$.
We see that at any given mass $m'$ there can be both upper and lower bounds on the mixing angle $\ep$; thus, there are two critical values of the mixing angle, which we call $\epr$ and $\etr$. Below $\epr$, dark photons are so weakly coupled that they cannot be produced readily enough to affect the evolution of the proto-neutron star. However, not all mixing angles $\ep>\epr$ are ruled out: at sufficiently large mixing angles, above $\etr$, dark photons are trapped before they exit the star, allowing energy to be efficiently reprocessed in the form of thermal neutrinos. Thus, only in the range $\epr \leq \ep \leq \etr$ can the dark photons cause a violation of the bound in \Eq{lum-bound}.

The dark-photon luminosity is given by a volume integral of the differential power, $dP/d\omega dV$. Before focusing on the dark photon, we 
describe the luminosity for an arbitrary differential power. Of relevance is the luminosity in new particles that cannot be reprocessed efficiently as neutrino energy. We are thus interested in the luminosity {\it from} the zone of neutrino diffusion {\it to} a zone in which neutrinos are not produced efficiently. In other words, we want to find the energy that escapes from behind the ``neutrinosphere'' (the isotherm where Standard-Model neutrinos approximately transition from diffusion to free streaming) defined by some radius $R_\nu$. In general, this may be written
\beq \label{lum-m-ep-R}
L(m',\ep,R_\nu,R_{\rm far}) = \int\limits_{r=0}^{R_\nu} \int\limits_{\omega=m'}^{\infty} \exp \bL - \tau(m',\ep,\omega,r,R_{\rm far}) \bR \frac{dP(m',\ep,\omega,r)}{dV d\omega} d \omega dV,
\eeq
where $\tau$ is the optical depth of a dark photon produced at radius $r$ and the energy integral starts at\footnote{The gravitational potential well of the proto-neutron star can in principle trap dark photons with small boosts \cite{Dreiner:2003wh}. However, we find that at most only about ? O(10\%) of the power is gravitationally trapped for masses of interest, which is the same order of magnitude as several other effects we have ignored, so we will henceforth omit this effect.} $m'$. For a dark photon traveling radially outward, we have
\beq \label{tau-m-ep-R}
\tau(m',\ep,\omega,r,R_{\rm far}) = \int_r^{R_{\rm far}} \Gamma'_{\rm abs}(m',\omega,\ep, \tilde  r)  d \tilde r ,
\eeq
for an absorptive width $\Gamma'_{\rm abs}$; see \App{opt-depth-details} for the technical details of this calculation. 
As discussed below, $dP/dV d\omega$ is proportional to the absorptive width, $\Gamma'_{\rm abs}$ using the principle of detailed balance, 
so we can simply calculate $\Gamma'_{\rm abs}$ for any given interaction. 

Two radii appear in \Eq{lum-m-ep-R}. The radius $R_\nu \sim \cO(40\km)$ of the neutrinosphere is the radius outside of which most neutrinos free stream until arriving at Earth. 
For definiteness, we define $R_\nu$ to be the radius where the temperature of the star has fallen to $3\mev$. 
This is roughly consistent with the condition for neutrino free streaming. The radius $R_{\rm far} \sim \cO(100\km-1000\km)$ is some ``far radius'' outside of which neutrinos no longer are produced efficiently. The reason these radii enter our calculations in such a fundamental way is that they provide a measure of the amount of energy that is diverted {\it away from} neutrinos in a manner that can alter the observed neutrino signal. If we were interested in the ability of terrestrial experiments to detect the new physics particles, $R_{\rm far}$ would be the distance to the Earth, which takes into account interactions with the progenitor star material outside of $R_\nu$, interactions in the circumstellar medium, and the contribution from the column density of the interstellar medium across the remaining distance. However, the dark photons need not travel very far to have an effect on the evolution of the neutrino flux from the supernova explosion. Instead, the ``Raffelt criterion'' in \Eq{lum-bound} says that if energy is taken away from the core and deposited in a different region of the star, the energy is effectively lost because it becomes unavailable to neutrinos. This diversion depletes the fuel of the nuclear ``engine'' that allows the cooling timescale to be $t_{\rm cool} \sim \cO(10\sec)$.

\begin{table}[t]
\begin{center}
\begin{tabular}{|c|c|c|} \hline
Possible values for $R_{\rm far}$ & distance & physical justification  \\ \hline\hline
$R_{\rm gain}$ & 100 km & for $r>R_{\rm gain}$, $\nu$ capture exceeds $\nu$ production \\ \hline
$R_{\rm shock}$ & 1000 km & for $r>R_{\rm shock}$, material is not yet shock heated \\ \hline
\end{tabular}
\end{center}
\caption{We consider two choices for the distant radius $R_{\rm far}$ beyond which $A'$ particles must transport energy 
to affect the neutrino cooling phase.} 
\label{r-fars}
\end{table}%

There are a number of reasonable choices for $R_{\rm far}$; the only strict requirement is $R_{\rm far}>R_\nu$. If $R_{\rm far}$ is too close to $R_\nu$, we would erroneously conclude that arbitrarily well-mixed dark photons would carry away too much energy, because for dark photons produced in the final shell the integral in \Eq{tau-m-ep-R} would go over a very small range and $\tau \to0$. However, this is unphysical: if the energy from the dark photons can be reprocessed by Standard-Model particles into neutrino energy, the dark photons do not provide an important energy sink. For this reason, we suggest that the lower bound on $R_{\rm far}$ is the neutrino gain radius $R_{\rm gain} \sim \cO(100\km)$, outside of which neutrino production has a lower rate than neutrino absorption~\cite{Bethe:1992fq, Janka:2000bt}. A reasonable upper limit on $R_{\rm far}$ is the shock radius $R_s \sim \cO(1000\km)$, outside of which matter is as yet uncompressed~\cite{Raffelt:1996wa}. Here, we will parameterize our uncertainty on $R_{\rm far}$ by using the gain radius and the shock radius, i.e.~$R_{\rm far} = 100\km$ and $1000\km$, as representative values. We list these in \Tab{r-fars} for reference.

We point out here that our method of calculating $L$ differs from prior work: we do not split the calculation into free-streaming and trapped regimes. Instead, we allow $\tau$ to ``speak for itself'' and suppress the integration in parts of parameter space where the optical depth is large and the particles are mostly trapped. As we discuss in greater depth in \Sec{DFA-breakdown}, this has very important consequences at large mixing angles. 
At large mixing angles, we find that the energy spectrum is not thermal.  For a given value of $\ep$, assuming a thermal spectrum underestimates 
the luminosity and thus leads to weaker limits.

\subsection{Uncertainties Regarding the Explosion of SN1987A}
\label{stellar-unc}

SN1987A is a promising environment for examining new physics because of the combination of the unique physical conditions attained in the star and the proximity of the explosion. However, constraints on new physics from the observation of SN1987A are inherently limited by difficulties in understanding the detailed process of the supernova even in the minimal case with no new physics. Many aspects of SN1987A remain poorly understood, from the nature of the progenitor to the primary driver of the ``shock revival'' required to sustain the supernova explosion. The mass of the progenitor star is only bracketed within a factor of two, and consequently the temperature and density profiles have large, qualitative uncertainties.

Given the uncertainties in modeling SN1987A, it is sufficient to use the conservative limit on the luminosity in \Eq{lum-bound} to derive  
bounds.  
To aid our analytic understanding, we will refer throughout the text to a ``fiducial'' model, 
a simple analytic supernova profile as advocated for in~\cite{Raffelt:1996wa} 
\beq \label{fiducial}
\rho(r) = \rho_c \times \cbL \begin{array}{cc} 1+k_\rho(1-r/R_c) & r<R_c \\ (r/R_c)^{-\nu} & r \geq R_c \end{array} \right., \qquad 
T(r) = T_c \times \cbL \begin{array}{cc} 1+k_T(1-r/R_c) & r<R_c \\ (r/R_c)^{-\nu/3} & r \geq R_c \end{array} \right. .
\eeq
This model is described by a core radius, $R_c$, outside of which the temperature and density fall like power laws as functions of $r$. As fiducial parameters we will take $k_\rho = 0.2,$ $k_T = -0.5,$ $\nu=5,$ and  $R_c=10\km$ with core temperature and density of $T_c=30\mev$ and $\rho_c=3\tenx{14}\g/\cm^3$ and a uniform proton fraction of $Y=0.3$. Given these quantities, we also consistently solve for the electron chemical potential, as outlined in~\cite{Braaten:1993jw}, which will be important for scattering and decay processes involving electrons. 

\begin{figure}[t!]
\begin{center}
\includegraphics[width=\textwidth]{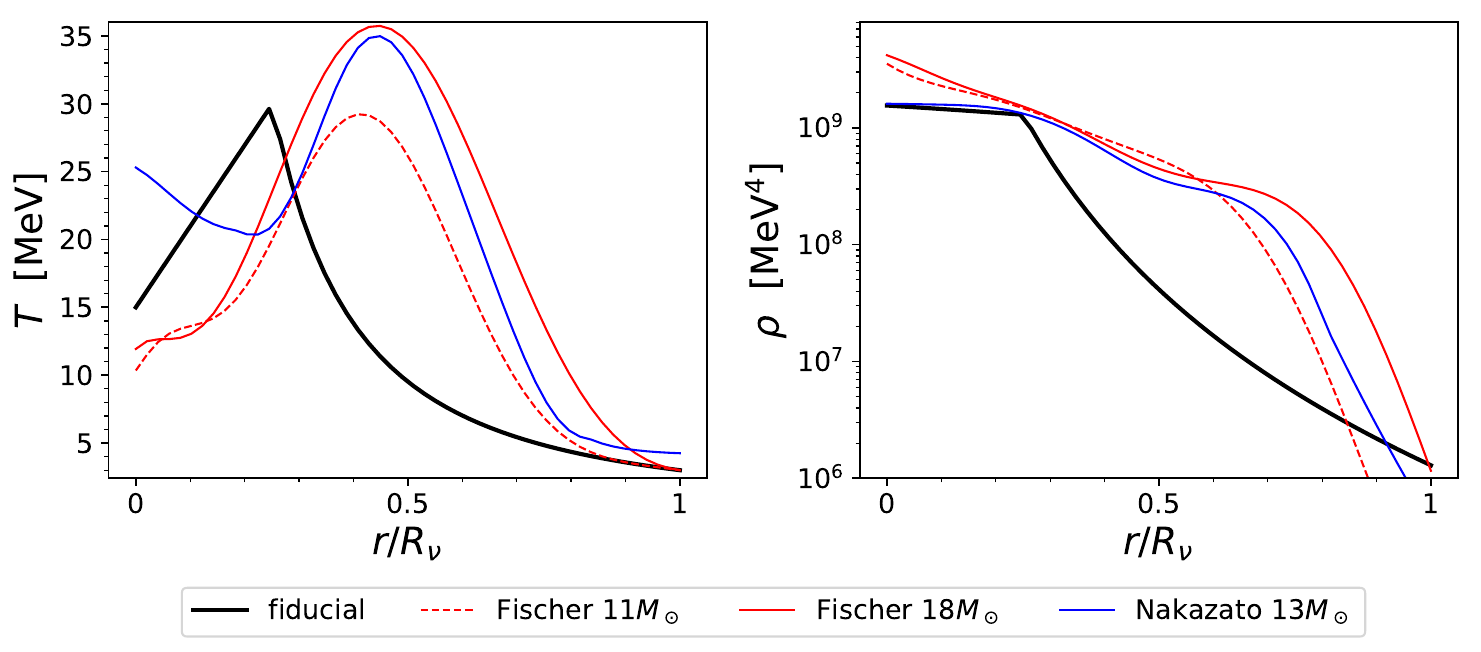}
\caption{Profiles of temperature and density for the different profiles described in \Sec{stellar-unc}: the ``fiducial model'' described in \Eq{fiducial}, two obtained from the reference runs of~\cite{Fischer:2016cyd}, and one profile from \cite{Nakazato, Nakazato:2012qf}. We cut off each plot at $R_\nu$, which 
we define as the radius at which the temperature is $T=3\mev$ for the profiles of \Eq{fiducial} and \cite{Fischer:2016cyd} and the radius which equals the thermal neutrino mean free path for the profile of \cite{Nakazato:2012qf}.
Note that the value of $R_\nu$ differs across the simulations, and in the figures each profile is scaled by its corresponding value for $R_\nu$.
}
\label{profile-plots}
\end{center}
\end{figure}

Our fiducial model, while convenient for analytic purposes, will inevitably mischaracterize the core collapse to some degree. To parameterize the uncertainties arising from the exact nature of the supernova explosion, we will also show bounds using numerical profiles from a variety of different sources~\cite{Nakazato, Nakazato:2012qf,Fischer:2016cyd}. The profile from~\cite{Nakazato, Nakazato:2012qf} is the 13 solar mass progenitor at $t=1$ sec post bounce and with a 100 msec shock revival time inserted by hand. This work includes a solution to a neutrino radiative hydrodynamical code before shock revival and a solution to the flux-limited diffusion equation after cooling has commenced \cite{Nakazato:2012qf}. We refer to this profile as ``Nakazato $13M_\odot$'' in what follows. The profiles presented in~\cite{Fischer:2016cyd} are of 11.8 or 18
solar-mass progenitors, taken at $t=1\sec$ after collapse. These solve the Boltzmann equation for neutrino transport with the AGILE-BOLTZTRAN code and an equation of state based on known nuclear isotopes and relativistic mean field models~\cite{Hempel:2009mc,Liebendoerfer:2000fw,Liebendoerfer:2001gu,Liebendoerfer:2002xn,Liebendoerfer:2003es}. We refer to these as ``Fischer $11M_\odot$'' and ``Fischer $18M_\odot$'', respectively. The fiducial model matches these profiles quite well in the interior of the star. We will emphasize the fiducial model in what follows because of its analytic nature and for ease of comparison with past work. However, this emphasis should not be interpreted as an endorsement of the fiducial profile over the numerical profiles we consider.  

We plot the temperature and density as functions of radius in \Fig{profile-plots} for the fiducial model and for the four numerical models, and we list some of their benchmark data in \Tab{radii-of-interest}. As is apparent, these differ by as much as an order of magnitude in certain parts of the proto-neutron star. Such uncertainties imply that any given profile does not capture all of the subtleties of the behavior of the nuclear matter in this violent environment. Indeed, constraints derived with any given profile are necessarily approximate\footnote{In fact, a neutrino burst that sustains an explosive front compatible with the observed energetics is challenging \cite{Janka:2012wk, Burrows:2012ew} (though not impossible \cite{Melson:2015spa, Sukhbold:2015wba}) to obtain in numerical simulations, and it is possible that the explosion may instead be revived by qualitatively different physics far from the core \cite{Blum:2016afe}. In this work we rely on the standard neutrino-induced explosion picture and refer the reader to \cite{Burrows:2012ew, Blum:2016afe} for more discussion.}: given the limited understanding we currently have of the core collapse mechanism and the nature of SN1987A, more observations of core collapse supernovae are necessary to determine whether any given point, especially one near the boundary of our excluded region, is in fact incompatible with observations. 

 
Despite these caveats, we will ultimately see that these ``boundary uncertainties'' affect a subdominant part of the parameter space. 
Moreover, the bounds derived from these various profiles provide an estimate of the systematic uncertainties.  
We will refer to the region excluded for all profiles as the ``robustly excluded'' region, which we believe takes into account the dominant 
systematic uncertainties of supernova astrophysics.

\begin{table}[t!]
\begin{center}
\begin{tabular}{|c|c|c|c|c|} \hline
 & fiducial & Fischer $11M_\odot$ & Fischer $18M_\odot$ & Nakazato $13M_\odot$ \\ \hline
$R_\nu$ [km] & 39.8   & 24.9  & 23.6 & 25.6 \\ \hline
$R_c$ [km] & 10 & $\sim 10$ & $\sim 11$  & $\sim 13$ \\ \hline
$\rho_c \ [10^{14} \g/\cm^3]$ & $3.0$ & $\sim 1.8$ & $\sim 1.2$ & $\sim 1.0$ \\ \hline
$T_c \ [\mev]$ & 30 & $\sim 29$ & $\sim 36 $ & $\sim 34 $ \\ \hline
\end{tabular}
\end{center}
\caption{Radii of interest for different profiles. We define the core radius $R_c$ as the radius of peak temperature. The maximum constrained mass $m'_{\rm max}$ in our final results in \Sec{sec:results} is essentially determined by the Boltzmann tail of the core temperature $m'_{\rm max} \sim 3 T_c$.}
\label{radii-of-interest}
\end{table}%

\section{Dark Photons in SN1987A}
\label{sec:A-prime-physics}

In this section, we discuss how to derive constraints from SN1987A on dark photons in the $m'-\ep$ plane.  
Two new elements in our calculation, compared to previous work, are that we include finite-temperature and finite-density 
effects~\cite{An:2013yfc} and we do not 
require that the emitted dark photons follow a Planck distribution, which is relevant for  $m' < 2 m_e$ and at large $\ep$.  We also include semi-Compton scattering, which is relatively important compared to inverse bremsstrahlung at high dark photon energies and in regions of low nuclear density.
These changes have important effects on the resulting constraints.

%

Limits that neglect finite-temperature and finite-density effects are accurate only for dark photon masses $m' \gg \omega_p$, where $\omega_p$ is the plasma mass of the Standard Model photon. For lower $m'$, the effects of finite temperature and density on the gauge-boson mixing angle cause the constraints to be systematically weakened compared to the results in \eg\cite{Rrapaj:2015wgs}, which assumed a constant (vacuum) value of the mixing angle at all values of $m'$ and all radii. Instead, the mixing angle in a medium, which we denote $\epm$, is given by \Eq{in-medium-angle} for each polarization separately.  The fact that $\epm$ differs for the longitudinal and transverse polarizations 
of the dark-photon field has important phenomenological consequences, as emphasized in~\cite{An:2013yfc}.

The main contribution to the polarization tensor of the photon, which determines the photon's plasma mass, comes from electrons,  
which are relativistic and degenerate for all radii within $R_\nu$, i.e.~$m_e \ll T \ll \mu_e$, where $\mu_e$ is the chemical potential.  
In this region we can use Eq.~(77) of~\cite{Braaten:1993jw} with the conventions of~\cite{An:2013yfc} to get the Standard-Model photon polarization tensor. As mentioned in Sec.~\ref{sec:intro}, the real part of this tensor is proportional to the plasma mass of the Standard Model photon,
\beq  \label{pol-ten}
\begin{array}{ll}
\re \Pi_L = \frac{3\omega_p^2}{v^2}  \pL 1-v^2 \pR \bL \frac1{2v} \ln \pL \frac{1+v}{1-v}  \pR -1 \bR, &\qquad\omega_p^2 = 4\pi \alpha_{\rm EM} n_e/E_F
\\  \re \Pi_T =\frac{3\omega_p^2}{2v^2} \bL 1 - \frac{1-v^2}{2 v} \ln \pL \frac{1+v}{1-v}  \pR \bR,  &\qquad E_F^2 = m_e^2 + (3\pi^2 n_e)^{2/3}.
\end{array}
\eeq
In \Eq{pol-ten} we define a photon velocity, $v$, via $| \vec k |^2 = \omega^2 v^2$, which can be small in medium. The imaginary part of the Standard-Model photon polarization function is proportional to the total width, $\im \Pi_{L,T} = - \omega \pL \Gamma_{{\rm abs}|L,T} - \Gamma_{{\rm prod}|L,T} \pR ,$ where the absorptive width $\Gamma_{\rm abs}$ is the total rate at which Standard-Model photons of energy $\omega$ are absorbed, and $\Gamma_{\rm prod}$ is the rate at which they are produced~\cite{An:2013yfc,Weldon:1983jn}. In local thermal equilibrium, these rates are related by $\Gamma_{\rm prod} = e^{-\omega/T}\Gamma_{\rm abs}$. Since Standard Model photons are in thermal equilibrium, the imaginary part of the photon polarization tensor is
\beq \label{impi}
\im \Pi_{L,T} (\omega,r) = - \omega \pL 1 - e^{-\omega/T} \pR \Gamma_{{\rm abs}|L,T}(\omega,r).
\eeq
Together, \Eqs{pol-ten}{impi} allow one to calculate the mixing angle under all conditions. 

The differential power in photons is related to the absorptive width by~\cite{An:2013yfc}
\beq \label{diff-power}
\frac{dP(r)}{ dV} = \int \frac{d^3 k}{(2\pi)^3} \omega \sum \Gamma_{{\rm prod}|L,T}(\omega,r)= \int \frac{\omega^3 v \,  d \omega}{2\pi^2} e^{-\omega/T} \sum_{\rm\,(in\,eq.)} S_{L,T} \Gamma_{{\rm abs}|L,T}(\omega,r)  ,
\eeq
where $S_L(S_T)=1(2)$ is the spin degeneracy of a given polarization state of the Standard-Model photon. We are, of course, interested in the power carried by dark photons, which are exclusively produced through kinetic mixing with Standard-Model photons. The power $dP'$ of interest is obtained by making the substitution
\beq
\Gamma_{{\rm abs}|L,T}(\omega,r) \to \Gamma'_{{\rm abs}|L,T}(m',\ep,\omega,r) = (\epm)_{L,T}^2 \Gamma^{\rm(in\,eq.)}_{{\rm abs}|L,T}(\omega,r)|_{v = \sqrt{1-m'^2/\omega^2}}\,.
\eeq
Given the absorptive width for the Standard-Model photon, we may then use \Eq{diff-power} to calculate the differential power for dark photons of any mass and mixing angle. Since the optical depth $\tau$ is the integral of the absorptive width over all radii larger than $r$, as given in \Eq{tau-m-ep-R}, it follows that we need only to calculate $\Gamma_{\rm abs}$ to be able to find the luminosity in dark photons. 

We consider $A'$ production and absorption through bremsstrahlung in neutron-proton collisions, Compton-like scattering involving an $e^-$, and decay to $e^+e^-$ pairs (see Fig.~\ref{feyn-diags}). The absorptive width will generically be different for the longitudinal and transverse polarizations because the different polarizations have different dispersion relations in medium. We describe the calculations for these widths in detail in \App{app:mfp}; here we summarize the relevant results.

The dominant process in the core of the proto-neutron star is dark-photon bremsstrahlung, shown in the left panel of \Fig{feyn-diags}. The width for absorption through inverse bremsstrahlung is
\beq \label{Gamma-inv}
\Gamma_{{\rm ibr}|L,T}' =\frac{32}{3\pi} \frac{\alpha_{\rm EM} (\epm)_{L|T}^2 n_nn_p}{\omega^3} \pL\frac{\pi T}{m_N}\pR^{3/2} \langle \sigma_{np}^{(2)}(T) \rangle \times \bL  \frac{m'^2}{\omega^2} \bR_L,
\eeq
where 
\beq \label{eq:sigmanp}
\langle \sigma_{np}^{(2)}(T) \rangle = \frac12 \int_{-1}^1  d\cos\theta\, \int_0^\infty d x\,e^{-x} x^2  (1-\cos \theta)\,  \frac {d\sigma_{np}(x T)}{d\cos\theta}
\eeq
is the appropriately averaged neutron-proton dipole scattering cross section~\cite{Rrapaj:2015wgs}. The subscript on the final fraction of \Eq{Gamma-inv} indicates that it should be included for  $\Gamma_{{\rm br}|L}'$ only. As described in \App{app:mfp}, to the order of the soft radiation approximation that we can calculate~\cite{Nyman:1968jro,Rrapaj:2015wgs}, the rate for bremsstrahlung and inverse bremsstrahlung are equivalent. 
The reason is that this approximation is an expansion in $\omega/T_{\rm CM}$, where $T_{\rm CM}$ is the center-of-mass kinetic energy, so the dark photon energy should obey $\omega\ll T$, which formally corresponds to $e^{-\omega/T} \simeq 1$. 
However, if we do not enforce $\omega\ll T$, we discover large contributions from regions where the incoming nucleon energies are actually strongly Boltzmann suppressed.
Thus, to consistently account for larger values of $\omega$, we include a physical cutoff $\omega \leq T_{\rm CM}$, which allows a consistent reintroduction of a Boltzmann-like factor. With this substitution, the rate for bremsstrahlung production is no longer the same as the rate of inverse bremsstrahlung: we substitute $\langle \sigma_{np}^{(2)}(T) \rangle \to \langle \sigma_{np}^{(2)}(\omega, T) \rangle$, differing by changing the limits of integration of \Eq{eq:sigmanp} as $\int_0^\infty d x \to \int_{\omega/T}^\infty d x$.
For more details, see the discussion around \Eq{Gamma-prod-calc}. 

We refer to emission of a dark photon in photon-electron scattering as ``semi-Compton'' scattering; this is shown in the center panel of \Fig{feyn-diags}. Semi-Compton scattering is important for $m'<2m_e$ despite the fact that it is subdominant to bremsstrahlung in the core. It has typically been neglected in these models, which is a reasonable approximation assuming a thermal spectrum.  
However, it has a different scaling with energy and density than bremsstrahlung, and we find that semi-Compton scattering can 
dominate absorption far out in the star for $m'<2m_e$. Exact formulae for the different polarizations are cumbersome to write down due to the many mass and energy scales in the problem and the fact that the separate matrix elements are not Lorentz invariant.  
Nevertheless, we find an excellent fit to a complete calculation in the region $\omega \lesssim 200\mev$ and $r \leq R_\nu$ using the simple formula $\Gamma_{\rm sC} = \frac{8\pi \alpha^2 \epm^2 n_{e^-}}{3 E_F(r)^2} \sqrt{\frac{\omega_p(r)}\omega} \times \bL  \frac{m'^2}{\omega^2} \bR_L$  (in fact, we find that this approximation holds beyond $R_\nu$, but we only need it up to $R_\nu$). 
This may underestimate Pauli blocking for low dark-photon energies, but qualitatively reproduces the energy dependence from a more complete calculation.

A final contribution to the dark photon optical depth comes from the decay to $e^+e^-$, which is allowed for the dark photon but not the Standard Model photon~\cite{Braaten:1993jw}, and dominates the optical depth for $m' > 2m_e$. This has width
\beq \label{Gam-ee}
\Gamma'_{e^+e^-|L,T} = \Theta\!\pL m' - 2m_e \pR \frac{\alpha_{\rm EM} (\epm)_{L,T}^2 m'^2}{\sqrt{\omega^2-m'^2}} \int_{x_-}^{x_+} \frac{ {\sf m}(\omega x)^2_{L,T}\,  d x}{\exp\pL\frac{-x+\mu_e/\omega}{T/\omega}\pR+1} ,
\eeq
where the limits of integration are $x_\pm = \frac12\big[ 1\pm  \sqrt{\pL1- 4m_e^2/m'^2 \pR\pL1-m'^2/\omega^2\pR } \big],$ and the dimensionless matrix element squared ${\sf m}_{L,T}^2$ is given in \Eq{ee-matrix-element}. We emphasize that bremsstrahlung and inverse bremsstrahlung 
of photons are in equilibrium, $np \leftrightarrow np \gamma$, so that this process contributes both to dark-photon production and absorption. In contrast, positrons are very rare, so electron-positron coalescence is negligible, $e^+e^- \substack{\leftarrow \\[-.15em] \not \rightarrow} A'$, and this process only contributes to the absorptive width.

\subsection{Luminosity in the Low-Mixing Regime}
\label{dfa}

In this subsection, we consider small values of $\ep$, for which the optical depth is small.  
In this region of parameter space, there is a ``resonance'' in $\epm$.  From \Eq{in-medium-angle}, we see that a 
necessary condition for such a resonance is $\im \Pi \ll \re \Pi$.  Using \Eqm{pol-ten}{Gam-ee}, we verify that this condition is satisfied throughout the proto-neutron star.  
In the next section, we consider larger values of $\ep$ where there is no resonance.  

\begin{figure}[tbp]
\begin{center}
\includegraphics[width=.7\textwidth]{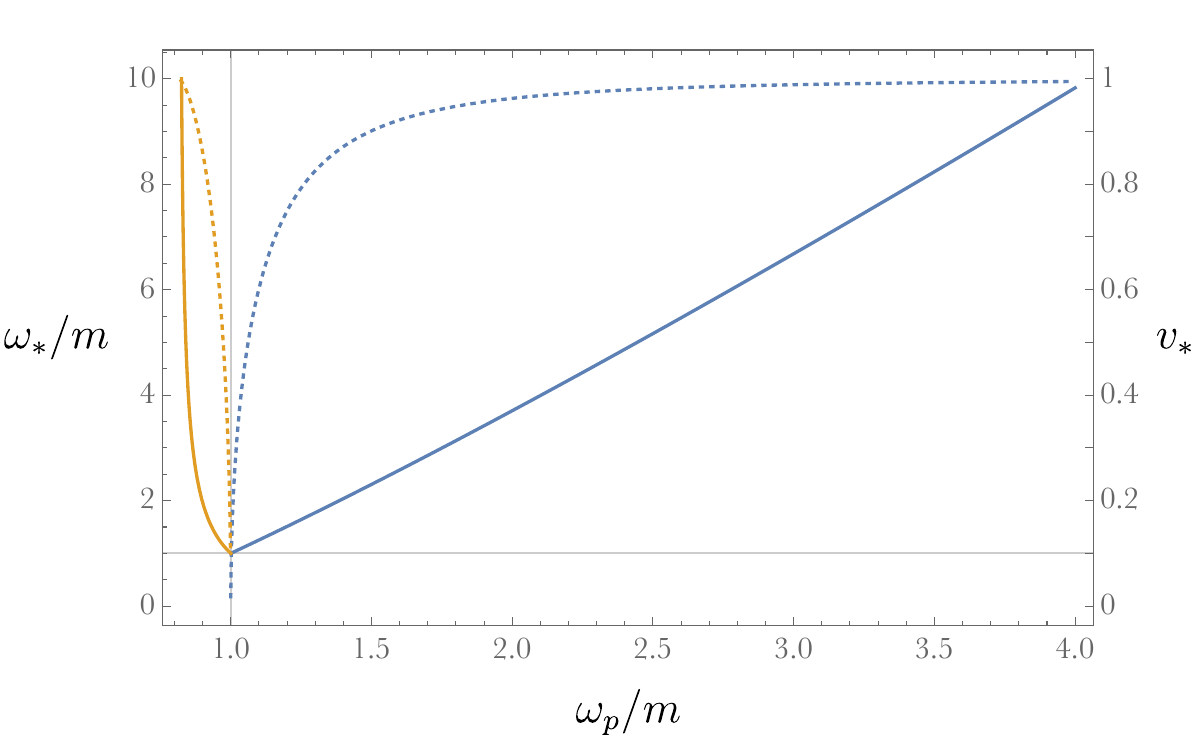}
\caption{Frequency $\omega_*$ (solid) and velocity $v_*$ (dotted) at which the dark photon hits a resonance; we show the longitudinal (transverse) mode in blue (orange). For the longitudinal (transverse) mode, the resonant modes are on shell only if $\omega_p\geq m'$ $\left(\sqrt{\frac23}m'\leq \omega_p \leq m'\right)$. 
}
\label{res-om}
\end{center}
\end{figure}

On resonance we have $\re \Pi = m'^2$, and the differential power is
\beq \label{res-dP}
\pt_V\! P_* \equiv  \left.\frac{dP_{L,T}(m',\ep,\omega,r)}{ dV } \mR_{\rm \delta \text - fn.} =  \int d\omega \frac{\ep^2 m'^2 \omega^3 v^3 \delta(\omega-\omega_{*|L,T})}{2\pi\pL e^{\omega/T}-1\pR} \cbL 2+\frac{m'^2}{\omega_{*|L,T}^2}-\frac{3\omega_p^2}{m'^2} \bL \frac{m'^2}{\omega_*^2} \bR_L \cbR^{-1} ,
\eeq
where the factor $\bL \frac{m'^2}{\omega_*^2} \bR_L$ applies only for the longitudinal mode, as in \Eq{Gamma-inv}. As expected, the details of the production process have fallen out. This follows from the scaling $\Delta \omega_* \sim \sum \Gamma_{\rm prod} \sim {\rm Im}\Pi$, where $\Delta \omega_*$ is the width of the resonance. The $\omega_{*|L,T}$ in \Eq{res-dP} are the resonance energies for a given $\omega_p$ and $m'$. These are found by solving $\re\Pi(\omega_*)=m'^2$ and restricting to the mass shell with $v=v_*=\sqrt{1-m'^2/\omega_*^2}$. We show $\omega_*$ as a function of $\omega_p/m'$ in \Fig{res-om}: for the longitudinal (transverse) mode, resonance occurs for $\omega_p > m'$ ($\sqrt{2/3}~ m' < \omega_p < m'$). In the limit $\omega_p \to m',$ the resonance energy for each polarization approaches the mass, $\omega_* \to m'$, and the velocity becomes small. In \Eq{res-dP}, we see the familiar small-$m'$ scaling of $m'^2 \ep^2$ for $L$ modes versus $m'^4 \ep^2$ for $T$ modes \cite{An:2013yfc}. Moreover, $L$ modes are resonantly produced throughout the star where $m'<\omega_p$, while $T$ modes are resonantly produced only if radii exist at which $\omega_p < m' < \sqrt{3/2} \omega_p$~\cite{Braaten:1993jw}. For example, for the fiducial model, only dark photons of mass $1.3 \mev \lesssim m' \lesssim 16.7 \mev$ can be resonantly produced with the transverse polarization.

Since the density decreases monotonically, we find that dark photons can be efficiently absorbed near the site of their production. The $\tilde r$ integral for the optical depth is resolved by a delta function just as the $\omega$ integral is for the power, giving
\beq \label{tau-star}
\tau_* \equiv \left. \tau_{L,T}(m',\ep, \omega, r)\mR_{\rm \delta \text - fn.} = \frac\pi2 \frac{\ep^2 m'^2 \omega_p^2/\omega_{*|L,T} }{\mL \frac d{d\tilde r} \omega_p^2 (\tilde r) \mR_{\tilde r=r}}\, \frac{ 1+ \frac{\Gamma_{e^+e^-|L,T}'}{\Gamma_{{\rm sC}|L,T}'+\Gamma_{{\rm ibr}|L,T}'}}{1-e^{-\omega_{*|L,T}/T}}.
\eeq
Since $e^+e^-$ coalescence is not in equilibrium we have $\im \Pi \not \propto \sum \Gamma_{\rm abs}$, and a ratio of widths appears in the numerator of \Eq{tau-star}. If the $A'$ has additional absorptive or decay channels, their widths will also appear in the numerator of this expression.

There can only be a resonance for $m' \leq  \sqrt{\frac32}\omega_p(r=0) \sim 15\mev$ (for larger $m'$, one can still produce enough dark photons 
off-resonance to violate the Raffelt criterion, see Sec.~\ref{DFA-breakdown}).  
We find that for $\ep \sim 3\tenx{-9}/(m'/\mev)$, the Raffelt criterion is violated, corresponding to $\epr(m')$ in \Fig{schematic-luminosity}. At larger mixing angles, the resonant production is suppressed by resonant absorption. Using only the delta-function approximation for $\tau$ and $P$, we find that the bounds are lifted for mixing angles in excess of $\ep \sim 10^{-7}/(m'/\mev)$, corresponding to $\etr(m')$ in \Fig{schematic-luminosity}. With this simple approximation, we are able to rule out the region
\beq \label{d-fn-bound}
\text{$\delta$-function only:\qquad}  3\tenx{-9} \lesssim \ep \, m'/\mev \lesssim 10^{-7}, ~~m'\lesssim 15\mev .
\eeq
This partial result already differs from earlier work using related techniques~\cite{Bjorken:2009mm, Dent:2012mx, Kazanas:2014mca, Rrapaj:2015wgs}. The difference in the magnitude of the lower bound above a few MeV compared to~\cite{Kazanas:2014mca, Rrapaj:2015wgs} is due to $\cO(10)$ differences in our calculation of the differential power. The more consequential difference in the {\it shape} of our bound, which has a linear dependence on mass, is a consequence of including finite-mass effects on the mixing angle.

The delta-function approximation, which gives our result in \Eq{d-fn-bound}, fails in a dramatic but predictable way for large $\ep$, which 
we discuss next.

\subsection{Luminosity in the High-Mixing Regime}
\label{DFA-breakdown}

The delta-function approximation is not always suitable for obtaining the luminosity emitted in dark photons. The approximation fails because two delta functions appear in the integrals of \Eq{lum-m-ep-R}: one arises in the optical depth $\tau$, and another in the power $dP/dVd\omega$. Critically, the exponential of $-\tau_*$ means that the delta function in the opacity {\it competes against} the delta function contained in $dP/dVd\omega$. 
For $\tau_*\gg1$, the differential luminosity on resonance may be suppressed compared to the differential luminosity at higher energies, even if the standard condition for a resonance, $\im \Pi \ll \re \Pi$, holds. In other words, exponentiating $\tau$ at large values of $\ep$ may ``depress'' the peak of the would-be delta function in $dP/dVd\omega$ while leaving contributions at other energies untouched. The end result is that we can put bounds above $\etr(m')$ from \Eq{d-fn-bound}. With a proper treatment of the absorption mechanism, we can even place bounds stronger than those from assuming that particles in the high-mixing regime are thermalized at some radius. We now discuss these effects in more depth.

At sufficiently small mixing angles where $\tau$ is small, $dL \simeq dP$ is a good approximation, and the area under the peak of the resonance well approximates the full integral: even though this peak is very narrow, the height of the peak at resonance exactly compensates for the small width, as discussed around \Eq{res-dP}. This fails when the optical depth at the delta function peak is non-negligible. When $\ep$ is large, the dark photons from the delta function peak are trapped due to their large mixing angle, and the differential power at $\omega_*$ is no longer large enough to overcome the small width of the resonance. 

\begin{figure}[tb]
\begin{center}
\includegraphics[width=0.54\textwidth]{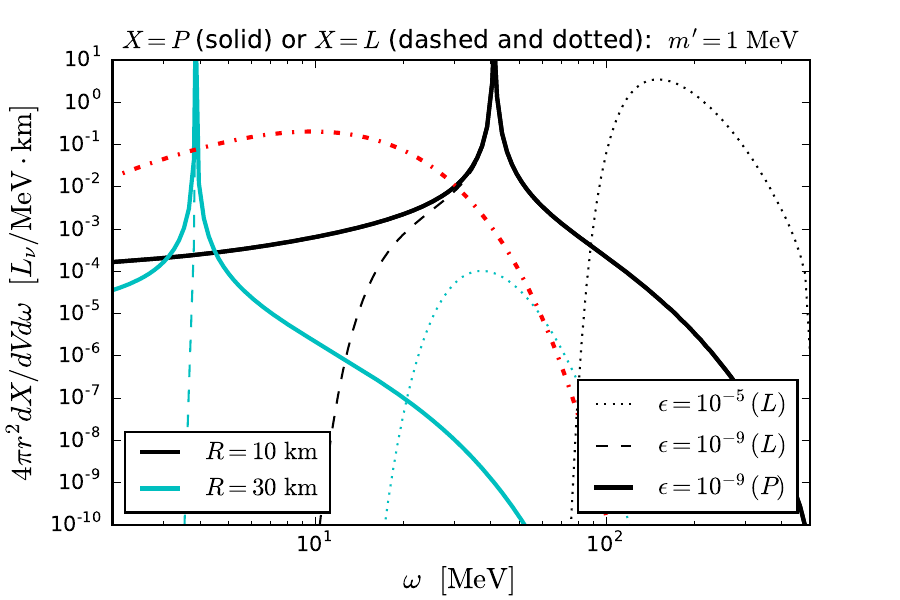}~~
\hskip-8mm 
\includegraphics[width=0.54\textwidth]{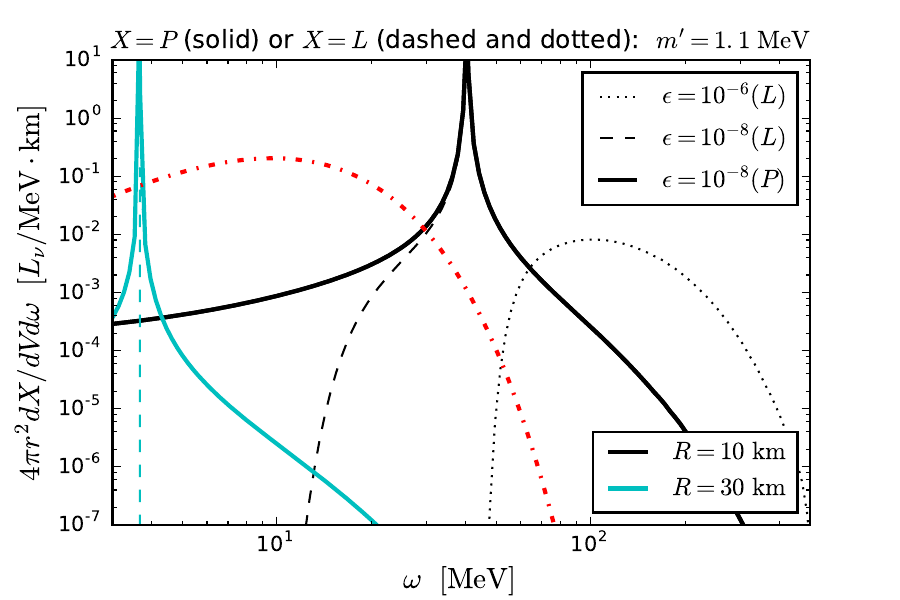}
\caption{ The differential power $dP/dV d\omega$ and luminosity $dL/dV d\omega$ for {\bf(left)} $m'=1\mev$ and {\bf(right)} $m'=1.1\mev$. The power is shown as the thick solid line. The luminosity is shown with a small (large) value of the mixing angle in dashed (dotted) lines: for $m'=1\mev$, small and large mixing corresponds to $\ep=10^{-9}$ and $10^{-5}$, respectively, while at $m'=1.1\mev$, small and large correspond to $\ep=10^{-8}$ and $10^{-6}$, respectively. The units are such that the integral of the curves together over $r$ and $\omega$ gives the luminosity of dark photons in units of $L_\nu$, so if this integral exceeds 1 the Raffelt bound is violated. The relative normalizations between curves demonstrate the effect of the overall $\ep$ scaling in $dP/dV d\omega$. We also compare to a Planck spectrum in the dot-dashed red line, shown at the radius $R_s$ for which the thermal luminosity $L_{\rm th}(m')= \pi^3 g_*(m') R_s^2 T(R_s)^4/30 = L_\nu$, following~\cite{Rrapaj:2015wgs}. At large mixing and low mass, the energy spectrum is neither resonant nor thermal: the peak of luminosity is above the Planck spectrum and above the resonance peak for large $\ep$ when $e^+e^-$ is not on shell.  This implies stronger limits than when simply 
assuming a thermal spectrum --- see \Sec{DFA-breakdown} for details.}
\label{non-res}
\end{center}
\end{figure}
We demonstrate this behavior in \Fig{non-res}, showing the case of dark photons with mass $m'=1$ and $1.1\mev$. These masses are chosen to be below and above the electron-positron pair production threshold, where the absorptive width scales differently. In this figure, we clearly see the breakdown of the delta function approximation in the large-mixing limit. At large mixing angles, the peak in $dP/dVd\omega$ is entirely depressed and contributes negligibly to the luminosity. The energy spectrum is instead peaked somewhere above $\omega_*$. This location and its height relative to the delta-function peak depends on the absorptive modes available to the dark photon. The energy that maximizes the differential luminosity depends on both $m'$ and $\ep$. At large $\ep$ and $m'<2m_e$, the energy spectrum of dark photons can have a peak of higher magnitude and larger central value than a Planck or delta-function spectrum. Thus, assuming a resonant or thermal energy spectrum at large mixing angle will {\it underestimate} the luminosity emitted in dark photons.

This conclusion regarding the energy spectrum of dark photon emission has important differences from standard methods in the ``trapping limit'' familiar from supernova bounds on sterile-neutrino production~\cite{Kainulainen:1990bn,Kuflik:2012sw}. Because sterile neutrinos scatter through weak processes with a rate that scales like $\Gamma_\nu \sim n s G_F^2$ and a mean optical depth that scales like $\tau_\nu \sim \omega^2$. At sufficiently large mixing angles the high-energy sterile neutrinos scatter many times to lower energies, and the ones that escape have approximately thermalized with the ambient matter at the radius at which the density and temperature are low enough for them to free stream. For this reason, it is accurate to say that the sterile neutrinos are emitted like a blackbody from some surface at large radius. However, if the dark photon scatters primarily through inverse bremsstrahlung, the optical depth goes like $\tau' \sim \omega^{-5}$ for small masses, and the optical depth {\it decreases} with $\omega$. Thus, high-energy particles are able to escape {\it without scattering}, providing a conduit for energy loss that is decoupled from local temperatures. The main consequence of this scaling is that low mass dark photons are emitted from within a sizable part of the star's volume, unlike the ``surface emission'' of sterile neutrinos.

For this reason, when the delta function approximation is inaccurate we need to integrate over the full range of dark photon energies to assess the luminosity in dark photons. This can be computationally expensive, especially if the resonance is subdominant but not negligible. However, because the optical depth decreases with energy, we find that a suitable approximation is to simply drop the low-energy tail:
\alg{ \label{appx-L}
\left. \frac{dL}{dV} \mR_{\rm large\,\ep} &\approx  e^{ - \tau_*} \, \pt_V\! P_*+  \int\limits_{\omega_*+4\Delta \omega_*}^{\infty}d \omega \frac{dP(m',\ep,\omega,r)}{dV d\omega} e^{- \tau(m',\ep,\omega,r,R_\nu)},
}
where $\Delta \omega_* = \frac{\pt_V \! P_*}{dP/dV d\omega|_{\omega=\omega_*}} = \frac{ \pi \omega_* v_*^2}{ 2+ \pL m^2-3 \omega_p^2 \pR /\omega_*^2} \frac{\im \Pi}{m^2} $ is the width of the delta function peak, and the factor of 4 gives $>99\%$ containment of a Breit-Wigner peak of these approximate dimensions. The first term in \Eq{appx-L} is a rescaled version of the delta function peak discussed in \Sec{dfa}: the optical depth suppresses the height of the delta function, but otherwise the resonance operates as before. When $\ep$ is large $\tau_*$ becomes considerable, and the second term will dominates the luminosity.

The consequence of this nonthermal energy spectrum at high $\ep$ is that we may constrain fairly large values of $\ep$. Assuming that highly mixed $A'$s are emitted with a thermal energy spectrum is bound to underestimate the luminosity in dark photons at masses below the $e^+e^-$ threshold. This is because for a fixed $\ep^2$ a fixed number of dark photons are produced. Assuming thermalization, we have $L' \sim \ep^2T^4 R_s^2$ for some ``$A'$-sphere'' radius $R_s$, while with our approach we find that $L' \sim \ep^2 \omega_s^3 m'^2 R_c^3$, where $\omega_s \gg T$ is the maximum of the differential luminosity. We find that this energy $\omega_s^4 \sim n_n n_p \langle \sigma_{np}^{(2)} \rangle$ may be quite high because it is not determined solely by the temperature, but rather by the energy scales of the absorptive processes. 
 Since the proto-neutron-star core is close to nuclear saturation (so the interparticle spacing is small and $n_n$ and $n_p$ are large), 
the emission of high-energy $A'$ can overcome the Boltzmann suppression $e^{-\omega_s/T}$ and provide enough energy loss to violate the Raffelt criterion. 
While these arguments work well for $m'<2m_e$, the decay $A' \to e^+e^-$ invalidates them 
for $m'>2m_e$ and keeps $L'$ close to thermal.  This can be seen in \Fig{non-res}.

\section{Results}
\label{sec:results}

By integrating the exact expressions for $dL/dV d\omega$, we can determine whether a point in the $\ep-m'$ parameter space leads to 
a dark-photon luminosity that exceeds the luminosity in neutrinos. When the optical depth on resonance is $\tau_* \ll 1$, this integral is well approximated by the resonant power. When the optical depth at the resonance energy is $\tau_* \gtrsim \cO(1)$, we must integrate \Eq{appx-L} over $0 \leq r \leq R_\nu$ to find the luminosity that is drained from inside the neutrinosphere.

Despite our quantitative understanding of this procedure as described in detail above, several major, qualitative sources of uncertainty linger in the calculation. We are unable to evade order-of-magnitude uncertainties regarding: the exact nature of the progenitor star, including its mass and peak temperature; the correct location for $R_{\rm far}$, beyond which energy is no longer available to be reprocessed for the neutrino cooling phase; and the form of the high-energy cutoff for $\Gamma_{\rm br}$, which, because we work in the soft-radiation approximation, is not given simply by a Boltzmann factor. Subdominant sources of uncertainty from the soft radiation approximation, such as Pauli blocking of initial state nucleons, relativistic corrections to the Boltzmann distribution, and low-energy effects like suppression of bremsstrahlung from the Landau-Pomeranchuk-Migdal mechanism~\cite{Baier:1996vi} should also contribute corrections of order tens of percent to our calculation. More aggressive bounds, obtained in principle by limiting the amount of energy {\it transport} at the core radius, are subject to larger uncertainties due to the difficulty of modeling supernova explosions, so we do not pursue such bounds here. 
Irrespective of the uncertainties, we advocate for including finite-temperature effects with new particles in future simulations of core-collapse supernovae.  
Moreover, the assumption that the emitted particles (at large mixing) must follow a thermal distribution is not necessarily correct, as 
demonstrated by the dark-photon model, where the trapping probability decreases for larger energies.  
This would also be interesting to include in future studies.

\begin{figure}[t]
\begin{center}
\includegraphics[width=.5\textwidth]{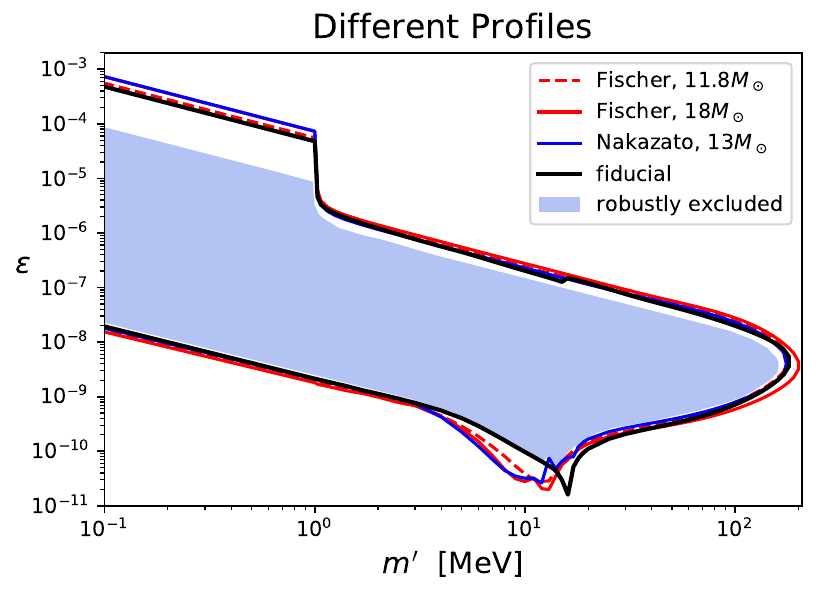}~~
\includegraphics[width=.5\textwidth]{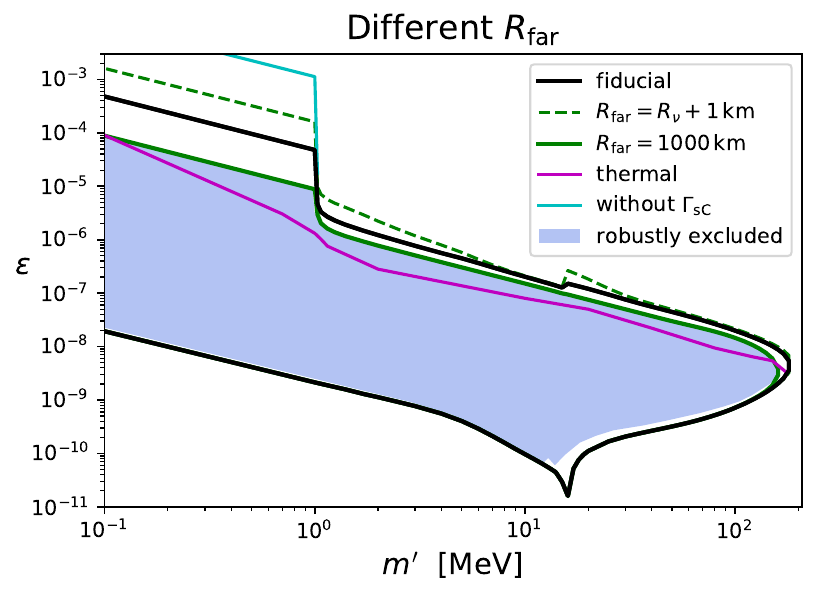}
\caption{Depiction of the dominant systematic uncertainties. Our ``fiducial'' profile as described in \Sec{stellar-unc} with $R_{\rm far}=100\km \simeq R_{\rm gain}$ is the thick black line. In the {\bf left} panel we vary the progenitor profile, while in the {\bf right} panel we vary the value of $R_{\rm far}$. In the right panel we also show some comparisons which are meant to illustrate the various magnitudes of the new physics we have accounted for: including finite temperature effects on $\epm$ but not accounting for the fully nonthermal $A'$ spectrum gives the magenta line; including all effects except for semi-Compton scattering gives the blue line.}
\label{syst}
\end{center}
\end{figure}

In \Fig{syst}, we estimate the size of the systematic uncertainties by plotting the excluded regions for all profiles and values of $R_{\rm far}$ that were discussed in \Sec{stellar-unc}. We emphasize that the entire lower boundary is fairly robust against the uncertainties we consider. Uncertainties regarding the temperature profile of the progenitor contribute to significant uncertainty on the location of the maximum constrained mass, however, and also on the upper boundary where $m'<2m_e$. The uncertainty in $R_{\rm far}$ is the greatest contribution to uncertainties for the upper boundary where $m' >2m_e$. This is because that boundary is mostly set by the decay length to $e^+e^-$, and the absorptive width scales linearly with $R_{\rm far}$. In contrast, for $m'<2m_e$, the dark photon is stable against decays,\footnote{The main decay channel is $A' \to \gamma \gamma \gamma$, with mean free path $\lambda_{\rm mfp} \sim5000\km (\ep/0.01)^{-2}(m'/\mev)^{-9}$~\cite{Pospelov:2008jk}.} and the uncertainty is driven by the details of the stellar core rather than by the physics near the neutrinosphere.
\begin{figure}[bt]
\begin{center}
\includegraphics[width=\textwidth]{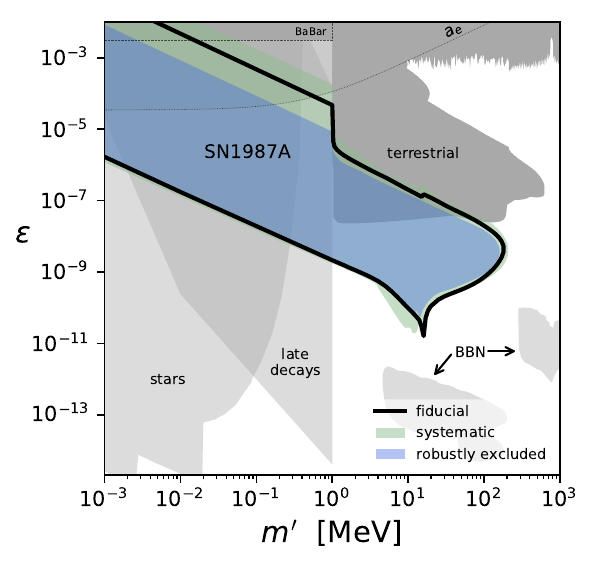}
\caption{Systematic uncertainties (green region) encompassing our ``robustly excluded zone'' (blue). The true boundary likely lies somewhere in the green region; we show our fiducial profile as the dotted black line. The blue area is excluded regardless of the perturbations we make to the physical inputs. We compare to bounds from other stars~\cite{An:2013yfc}, decays to three photons on cosmological timescales~\cite{Redondo:2008ec,Essig:2013goa}, and beam dumps, meson decays, and other terrestrial experiments~\cite{Alexander:2016aln}. The comparison of the electron anomalous magnetic moment in two different systems is shown in the hatched region, which has not previously been shown in this mass range.}
\label{final}
\end{center}
\end{figure}
The combination of the uncertainties in profile and $R_{\rm far}$ gives roughly up to an order of magnitude uncertainty at all points on the boundary. We can consider these uncertainties to be systematic error bars about the fiducial result. In \Fig{final}, we display the fiducial result as a red line, and the envelope of other bounds provide the extent of the green region. The blue region is constrained regardless of the perturbations we have described above, and we refer to it as ``robustly excluded''.

Compared to other results on the dark photon model~\cite{Dent:2012mx,Kazanas:2014mca,Rrapaj:2015wgs}, our constrained region is somewhat smaller (see \App{app:compare} for a direct comparison). The biggest change in our methodology is the inclusion of finite-temperature effects on the photon self-energy, which changes the asymptotic scaling at low mass to be constant in the product $\ep^2m'^2$ (rather than $m'$-independent). This has been accounted for in limits on dark photons from other stars~\cite{An:2013yfc}, but our results are the first to include plasma effects on the constraints from SN1987A. Another important change is that we discard the assumption that $A'$ particles thermalize at high mixing angle. We give a sense for the difference this makes by also showing the upper boundary of the constrained region using the (incorrect) assumption of thermalization. For masses asymptotically smaller than $2m_e$, the difference is of order a few. The discrepancy increases around a few hundred keV as the thermal production falls off while the resonant production remains essentially fixed. Above $2m_e$, where decays to electron-positron pairs dominate the absorptive width (and consequently cut off the high-energy tail of the differential luminosity), the assumption of thermalization leads to an $\cO(50\%)$ error in the bound. Finally, due to $\sim \cO(10)$ differences in our calculation of the bremsstrahlung width, our lower boundary moves up by a factor of $\sim\cO(3)$ compared to the results in~\cite{Rrapaj:2015wgs}. This lowers the maximum constrained mass because the highest constrained mass is determined by where the upper and lower boundaries (which are relatively steep functions of $m'$) intersect.

In \Fig{final} we also compare to other bounds for $m' \sim 1\kev-1\gev$. Dark photons below $\cO(100\kev)$ can disrupt the energetics of the Sun and other main sequence stars~\cite{An:2013yfc} (labelled ``stars''). Decays of $A' \to \gamma \gamma \gamma$, which are not observed from the time of big bang nucleosynthesis until the present day, provide the constraints in the region labelled ``late decays'': at the present cosmological epoch these decays would be observed as an isotropic diffuse X-ray background, and an injection of energy at the time of Big-Bang Nucleosynthesis is prohibited by the success of standard predictions. Using the width calculated in the Euler-Heisenberg effective theory~\cite{Pospelov:2008jk} and the model-independent bounds of diffuse X-ray background in~\cite{Essig:2013goa} for a decay to two photons, which we multiply by $3/2$ to account for the increased multiplicity in this model and rescale by the abundance from resonant production in the early Universe~\cite{Redondo:2008ec}, we obtain the lower bound of this region. Requiring $\tau \leq 1 \sec$ so that decays do not interrupt nucleosynthesis gives the upper line of this gray region~\cite{Redondo:2008ec}. We caution that threshold effects near $m' \lesssim 2m_e$, analogous to the effect of the $W,Z,$ and $t$ masses on the Higgs width to two photons, should in general {\it increase} this partial width beyond the rate calculated in the effective theory. This would open up the high-$\ep$ range of this region near $m' = 2m_e$ and deserves a more thorough investigation. 

We also show bounds on ``very dark photons'' from~\cite{Fradette:2014sza, Berger:2016vxi}.  These constraints are on isolated islands of $\ep$ as low as $\ep \gtrsim \few\tenx{-18}$, whose shapes are determined by considering precise channels of energy deposition from decays $A' \to e^+e^-$ during Big-Bang nucleosynthesis. 
Above $2m_e$, bounds from beam-dumps, $B$- and $\Phi$-factories, fixed-target experiments, and 
precision measurements of the muon and electron (``$a_e$'') anomalous magnetic moment are 
applicable~\cite{Bjorken:2009mm,Bjorken:1988as,Riordan:1987aw,Bross:1989mp,Batell:2009yf,Blumlein:2011mv,Andreas:2012mt,Pospelov:2008zw,Reece:2009un,Aubert:2009cp,Archilli:2011zc,Abrahamyan:2011gv,Merkel:2011ze,Merkel:2014avp,Babusci:2012cr,Babusci:2014sta,Lees:2014xha,Batley:2015lha,::2016lwm,Babusci:2015zda,Alexander:2016aln}.  
We have extended the bound on $a_e$ also below $2m_e$. 
The region labelled ``BaBar'' is a constraint obtained from a BaBar search for $e^+e^-\to \gamma + {\rm invisible}$, which applies 
for $m' < 2m_e$, where the $A'$ is long-lived and produced as $e^+e^-\to \gamma + A'$~\cite{Essig:2013vha}.  

\section{Conclusions}

We have examined the effects of dark photons on SN1987A, incorporating for the first time the effects of the finite temperature and density in this environment on the kinetic mixing angle. Resonant production is important at low mixing. These effects cause the constraints to lift at small $m'$ in such a way that the product $\ep m'$ is constant. Our result for the lower edge of excluded region is that we roughly require $\ep^2 (m'/\mev)^2 \lesssim 3 \tenx{-9}$ for $m'\lesssim 15\mev$, and $\ep \lesssim 10^{-9}$ for $15 \mev \lesssim m' \lesssim 120\mev$, when the production begins to be Boltzmann suppressed.

We have also carefully considered the non-thermal part of the dark-photon energy spectrum, discovering that the limits at very large mixing angles are somewhat more constraining than previously understood. The reason this is possible is that the energy spectrum at large mixing angle is not close to thermal. In fact, below the $e^+e^-$ threshold the peak of the differential luminosity is up to an order of magnitude higher than the temperature, and our bounds are a factor of a few stronger than they would be if we assumed that the spectrum were thermal.
This may have important implications for constraints on other dark-sector particles.  
We leave an investigation of this and variations of the simple dark-photon model to an upcoming publication~\cite{future-DM-paper}.

\section*{Acknowledgements}
We thank Kfir Blum, Alan Calder, Eder Izaguirre, Marc Kamionkowski, Georg Raffelt, Sanjay Reddy, Javier Redondo, and Natalia Toro for discussions, and we thank Alex Friedland, Ken'ichiro Nakazato, and Doug Swesty for their insights into the environment of the proto-neutron star.  
R.E.~is supported by the DoE Early Career research program DESC0008061 and through a Sloan Foundation Research Fellowship. 
SDM is supported by NSF PHY1316617.

\appendix
\section{Calculations for Production and Absorption Rates}\label{app:mfp}

In this appendix we discuss several details of the calculations for rates of production and absorption of dark photons. Following~\cite{Weldon:1983jn}, we define 
\beq \label{mfp-definition-0}
 \Gamma_{\rm abs} (\omega) = \frac1{4\pi} \frac1{2\omega} \int d\Omega' \,\cC[f_A]\,,
\eeq
where $K^\mu=(\omega, \vec k)$ is the photon four-momentum, $\Omega'$ is the photon scattering angle,
\alg{
\cC = \int\frac{d^3\vec p_1}{2E_1} \frac{f_p(E_1)}{(2\pi)^3} \int\frac{d^3\vec p_2}{2E_2} &\frac{f_n(E_2)}{(2\pi)^3}  \int\frac{d^3\vec p_3}{2E_3} \frac{1-f_p(E_3)}{(2\pi)^3} \int\frac{d^3\vec p_4}{2E_4} \frac{1-f_n(E_4)}{(2\pi)^3}   \times
\\& \qquad \times (2\pi)^4 \delta^4\pL K^\mu+ P_1^\mu +P_2^\mu -P_3^\mu -P_4^\mu \pR \mL \cM \mR_{\gamma np}^2
}
is the collision operator for the dark photon in the proto-neutron star, $P_i^\mu = (E_i,\vec p_i)$ are the nucleon four-momenta, $f_{n,p}$ are the nucleon distribution functions, discussed in detail below, and $\mL \cM \mR_{\gamma np}^2$ is the squared matrix element summed, not averaged, over all spins. The absorptive width defined in \Eq{mfp-definition-0} enters the polarization tensor and the renormalized mixing parameter as discussed in \Sec{sec:A-prime-physics}. We now calculate this width for the processes shown in \Fig{feyn-diags}.

\subsection{(Inverse) Bremsstrahlung}
In this subsection, we detail the calculation of the rates for bremsstrahlung production and absorption of the Standard Model photon in the environment of the proto-neutron star. The extension to dark photons is addressed at the end.

We will use the soft radiation approximation~\cite{Rrapaj:2015wgs,Nyman:1968jro} to calculate the rate for inverse bremsstrahlung. The soft radiation approximation is strictly appropriate only when photon energies are much below the other energy scales in the problem, $\omega \ll T, T_{\rm CM}$. However, this approximation is observed to reproduce data for energies in mild violation of the bound, $\omega \sim T\times \few$~\cite{Rrapaj:2015wgs}, so that we can use it for all of the parameter space. 

With this approximation, the same calculation is formally appropriate for the production and absorption rates of dark photons: although production and absorption are in equilibrium as discussed in \Sec{sec:A-prime-physics}, and thus are exactly related by the Boltzmann factor $e^{-\omega/T}$, in the soft-radiation approximation we keep only the leading-order contribution, and this factor becomes unity. In contrast, errors incurred in constructing the matrix element are $\sim \cO(|\vec p|/m_N)$, which scale like $|\vec p| \sim \sqrt{\omega/m_N} \gg \omega/m_N$. Therefore, corrections to the matrix element itself are of higher order than the Boltzmann factor. Nevertheless, we will find that a cutoff  that parametrically scales like the Boltzmann factor necessarily arises with the requirement that $\omega$ not exceed the energy of the collision: without specifying a shape or a new energy scale, ensuring that collisions not violate energy conservation reintroduces a high energy cutoff.

In the soft-radiation approximation, using Lorentz invariance, gauge invariance, and the fact that the photon couples to the dipole current in $n-p$ scattering, we find that the matrix element of interest\footnote{The photon couples to the quadrupole current in $p-p$ scattering~\cite{Nyman:1968jro}, which is suppressed by an additional factor of the kinetic energy divided by the total center of mass energy~\cite{Rrapaj:2015wgs}.} is~\cite{Nyman:1968jro}
\beq
\mL \cM \mR_{\gamma np}^2 = 4\pi \alpha_{\rm EM} \pL \varepsilon \cdot L \pR^2 \mL \cM \mR_{np}^2 \implies \oL{\mL \cM \mR}_{\gamma np}^2 = \frac{4\pi \alpha_{\rm EM}}{S_{L,T}} \sum_{\rm pols.}  \pL \varepsilon_\mu L^\mu \pR^2 \oL{\mL \cM \mR}_{np}^2\,,
\eeq
where $4\pi \alpha_{\rm EM}=e^2$ is the $U(1)_{\rm EM}$ coupling; $\varepsilon_\mu$ is the photon polarization; the dipole current is $L^\mu=P_1^\mu/P_1 \cdot K - P_3^\mu/P_3 \cdot K$; $S_{L,T}$ is the spin degeneracy for the polarizations of the Standard Model photon, with $S_L=1$ and $S_T=2$; and $\oL{\mL \cM \mR}_{np}$ is the spin-averaged matrix element for neutron-proton scattering, discussed in~\cite{Rrapaj:2015wgs}. We point out that the longitudinal and transverse modes of the dark photon have different matrix elements, since they have 
different dispersion relations in medium.

The mean free path for bremsstrahlung using \Eq{mfp-definition-0} is now written
\alg{ \label{mfp-definition}
&\Gamma_{\rm SRA} =  \frac{ \alpha_{\rm EM} }{2\omega S_{L,T}} \int\frac{d^3\vec p_1}{2E_1} \frac{g_p f_p(E_1)}{(2\pi)^3} \int\frac{d^3\vec p_2}{2E_2} \frac{g_n f_n(E_2)}{(2\pi)^3}  \times
\\&  \times \int\frac{d^3\vec p_3}{2E_3} \frac{1-f_p(E_3)}{(2\pi)^3}   \int\frac{d^3\vec p_4}{2E_4} \frac{1-f_n(E_4)}{(2\pi)^3}  (2\pi)^4 \delta^4 \! \pL P_1^\mu+ P_2^\mu- P_3^\mu - P_4^\mu \pR  \oL{\pL \varepsilon \cdot L \pR}^2 \oL{\mL \cM \mR}_{np}^2,
}
where $ \oL{\pL \varepsilon \cdot L \pR}^2 \equiv \int d\Omega' \sum_{\rm pols.} \pL \varepsilon_\mu L^\mu \pR^2$. Because we work in the soft-radiation approximation, we drop $K^\mu$ in the delta function in \Eq{mfp-definition}, which formally eliminates the distinction between bremsstrahlung and inverse bremsstrahlung.

We make some simplifications to \Eq{mfp-definition}. First, we will assume that the nucleons are neither degenerate nor relativistic. This means that Pauli blocking is negligible: $1-f_p(E_3)\approx 1-f_n(E_4) \approx 1$. With this assumption, the second line of \Eq{mfp-definition} is Lorentz invariant, so we may evaluate the second line in any frame that we choose. We choose the center-of-mass frame, $E_3=E_4=m_N + T_{\rm CM}/2 = E_{\rm CM}/2$  (where $m_N$ is the nucleon mass), and we take this opportunity to define the center-of-mass kinetic energy $T_{\rm CM}=(\vec p_1 -\vec p_2)^2/4m_N = E_{\rm CM}-2m_N$ for future convenience. We also recall that for elastic scattering
\beq
\frac{d \sigma_{np}}{d \Omega_{\rm CM}} = \frac{\oL{\mL \cM \mR}_{np}^2}{64\pi^2  E_{\rm CM}^2} \implies \oL{\mL \cM \mR}_{np}^2 =64\pi^2 E_{\rm CM}^2 \frac{d \sigma_{np}}{d \Omega_{\rm CM}},
\eeq
where $d \sigma_{np}/d \Omega_{\rm CM}$ is measured in the center-of-mass frame. (We point out here that we differ from~\cite{Rrapaj:2015wgs} in a factor of $v_{\rm rel}$.) In this frame the second line of \Eq{mfp-definition} is now
\alg{ \label{ltz-invt}
\int& \frac{d^3\vec p_3d^3\vec p_4}{(2\pi)^2 E_{\rm CM}^2} \delta^4 \!\pL P_1^\mu +P_2^\mu -P_3^\mu -P_4^\mu \pR\oL{\pL \varepsilon \cdot L \pR}^2 \oL{\mL \cM \mR}_{np}^2  =
\\ & \qquad\qquad\qquad \qquad= 16 \int d\Omega_{\rm CM} d |\vec p_3| |\vec p_3|^2 \oL{\pL \varepsilon \cdot L \pR}^2 \delta(E_{\rm CM}-E_3-E_4) \frac{d \sigma_{np}}{d \Omega_{\rm CM}}\,, }
where the energy delta function is $\delta(E_{\rm CM}-E_3-E_4) = \frac{m_N}{ 2  p_* } \delta\!\pL |\vec p_3|- p_* \pR $, which is satisfied at $p_* = \sqrt{m_N T_{\rm CM} }$, and the angle that $\vec p_3$ makes with respect to $\vec p_1$, $d\Omega_3$, is the scattering angle of the incoming $p_1$ evaluated in the center of momentum frame, $d\Omega_{\rm CM}$.

Doing the integral over $d |\vec p_3|$ results in an overall factor $m_N^{3/2}\sqrt{T_{\rm CM}}/2$. We plug this factor and the result of \Eq{ltz-invt} back into \Eq{mfp-definition} to get
\alg{ \label{mfp-simplify-2}
\Gamma_{\rm SRA} &=\frac{4\alpha_{\rm EM} m_N^{3/2}}{(2\pi)^6  S_{L,T}\omega}  \int  \frac{d^3 p_1 f_p(E_1)}{E_1}  \frac{d^3 p_2 f_n(E_2)}{E_2} d \Omega_{\rm CM} \oL{\pL \varepsilon \cdot L \pR}^2 \sqrt{T_{\rm CM}}  \frac{d \sigma_{np}}{d \Omega_{\rm CM}}.
}
We now use explicit forms of the distribution functions. For the non-degenerate nucleons, we take a rescaled Maxwell-Boltzmann distribution function, $f_{n,p}(E_i) = c e^{-E_i/T} $. Here, $c$ is defined such that $g_{n,p} \int \frac{d^3p_i}{(2\pi)^3} f_{n,p}(E_i) = n_{n,p},$ where $n_{n,p}$ is the observed (or simulated) number density of the nucleon. Because there are no anti-nucleons in the proto-neutron star, $g_{n,p}=2$ and explicitly
\beq \label{dist-func}
f_{n,p}(p_i) = \frac{n_{n,p}}2 \pL \frac{2\pi}{m_NT}\pR^{3/2} e^{-|\vec p_i|^2/2m_NT}.
\eeq
Plugging \Eq{dist-func} into \Eq{mfp-simplify-2} and simplifying, we have
\beq
\Gamma_{\rm SRA}  = \frac{ \alpha_{\rm EM}   n_nn_p}{8 \pi^3 S_{L,T} \omega T^3  m_N^{3/2}} \int d \cos \theta_{\rm CM} \frac{ d^3\vec p_1}{E_1}\frac{ d^3\vec p_2}{ E_2} \oL{\pL \varepsilon \cdot L \pR}^2 e^{-\frac{|\vec p_1|^2+|\vec p_2|^2}{2m_NT} }  \sqrt{T_{\rm CM}}  \frac{d \sigma_{np}}{d \cos \theta_{\rm CM}},
\eeq
where we have done the trivial integration over $d \phi_{\rm CM}$. We continue by changing coordinates to $\vec p_\pm = \vec p_1 \pm \vec p_2$, with $d^3\vec p_1 d^3\vec p_2 = \frac18 d^3\vec p_+ d^3\vec p_-$ and $|\vec p_1|^2+|\vec p_2|^2=(|\vec p_+|^2+|\vec p_-|^2)/2$: 
\alg{ \label{mfp-simplify-3}
\Gamma_{\rm SRA}  &= \frac{\alpha_{\rm EM}   n_nn_p}{64 \pi^3 S_{L,T} \omega T^3  m_N^{3/2}} \int d\cos \theta_{\rm CM}\frac{ d^3\vec p_+ }{E_1} \frac{d^3\vec p_-}{ E_2} \oL{\pL \varepsilon \cdot L \pR}^2e^{-\frac{|\vec p_+|^2+|\vec p_-|^2}{4m_NT} } \sqrt{T_{\rm CM}} \frac{d \sigma_{np}}{d\cos  \theta_{\rm CM}}.
}
This is not trivial to evaluate in generality, because $E_1$ and $E_2$ are functions of all six components of the incoming particle momenta. However, in the nonrelativistic limit we may take $E_1 E_2  \simeq m_N^2$. With this approximation, the width is
\alg{ \label{mfp-simplify-4}
\Gamma_{\rm SRA}  &\simeq \frac{\alpha_{\rm EM}   n_nn_p }{64 \pi^3 S_{L,T} \omega T^3  m_N^{7/2}} \int d\cos \theta_{\rm CM}   d^3\vec p_+ d^3\vec p_- \oL{\pL \varepsilon \cdot L \pR}^2  e^{-\frac{|\vec p_+|^2+|\vec p_-|^2}{4m_NT}  }  \sqrt{T_{\rm CM}} \frac{d \sigma_{np}}{d \cos \theta_{\rm CM}},
}
which is only a function of $|\vec p_+|$ and $|\vec p_-|$. Consequently, $ d^3\vec p_+ d^3\vec p_-  \to 16\pi^2 |\vec p_+|^2|\vec p_-|^2d |\vec p_+| d|\vec p_- |$.

Now we evaluate $\oL{\pL \varepsilon \cdot L \pR}^2$. In vacuum, we have  
\alg{\int  d\Omega'\sum_{\rm pols.}^{\rm(vac.)} (\varepsilon \cdot L)^2 = \frac{8\pi}3\frac{T_{\rm CM}}{m_N \omega^2} \pL 2 + \frac{\re\Pi}{\omega^2} \pR (1-\cos \theta)}
at leading order in $T_{\rm CM}$~\cite{Rrapaj:2015wgs}. However, at finite temperature the longitudinal and transverse polarizations of the 
Standard-Model photon are {\it not} the same, as they have different dispersion relations, and thus $\re \Pi_L \neq \re \Pi_T$. Rather than using the vacuum polarization sum, we evaluate the polarization sums in a relativistic medium. Here we have
\alg{ \label{projection-operators}
P^{L,T}_{\mu\nu}&\equiv \sum_{{\rm pols.}|L,T} \varepsilon_\mu \varepsilon_\nu = \cbL \begin{array}{ll}(1-\delta_{\mu0}) (1-\delta_{\nu0}) \pL \delta_{ij}- k_ik_j/ \vec k \cdot \vec k\pR
&~~ (T) \\  -g_{\mu \nu} + K_\mu K_\nu/ K^2 -P^T_{\mu\nu} &~~ (L)  \end{array} \right.,
}
with $\vec k \cdot \vec k \equiv \omega^2 v^2$, where the photon velocity $v^2$ need not be 1. To leading order, we find that the projected integrals of $L^2$ over the dark-photon scattering angle $d \Omega'$ are 
\alg{ \label{pols-separate}
\int  d\Omega' P^T_{\mu\nu} L^\mu L^\nu &= \int  d\Omega' \bL \vec L \cdot \vec L - \frac{(\vec L \cdot \vec k)^2}{\omega^2 v^2} \bR \simeq \frac{16\pi}3 \frac{T_{\rm CM}}{m_N \omega^2}(1-\cos \theta_{\rm CM}) ,
\\
\int \! d\Omega' P^L_{\mu\nu} L^\mu L^\nu &=  - \!\int \! d\Omega' \pL L \cdot L + P^T_{\mu\nu} L^\mu L^\nu \pR \simeq \frac{8\pi}3 \frac{T_{\rm CM}}{m_N \omega^2} \pL1-v^2\pR(1-\cos \theta_{\rm CM}).
}
Now we do the integral over $\theta_{\rm CM}$ by defining $ \sigma_{np}^{(2)}\equiv \int d \cos \theta_{\rm CM}(1-\cos \theta_{\rm CM})  \frac{d \sigma_{np}}{d \cos \theta_{\rm CM}}$, and the integral $\int |\vec p_+|^2 e^{-|\vec p_+|^2/4m_NT}d |\vec p_+| = 2\sqrt{\pi(m_N T)^3}$ is straightforward. With $S_T=2$ we have
\alg{
\Gamma_{{\rm SRA},T} &= \frac{4\sqrt\pi \alpha_{\rm EM} n_nn_p }{3 \omega^3 T^{3/2} m_N^3}  \int |\vec p_-|^2 d|\vec p_- | \exp\!\pL - \frac{|\vec p_-|^2}{4m_NT}  \pR  T_{\rm CM}^{3/2}\,  \sigma_{np}^{(2)}(T_{\rm CM}),
}
and $\Gamma_{{\rm SRA},L} = \pL1-v^2\pR \Gamma_{{\rm SRA},T}$. Finally, we change to a dimensionless variable $ x=T_{\rm CM}/T$ such that $ |\vec p_-|^2 = 4m_N T x$. The measure becomes $|\vec p_-|^2 d |\vec p_-| \to 4(Tm_N)^{3/2}  x^{1/2} d  x$, so 
\alg{
\Gamma_{{\rm SRA},T} &= \frac{16\alpha_{\rm EM} n_nn_p }{3\pi \omega^3} \pL \frac{\pi T}{m_N} \pR^{3/2}  \int_0^\infty d  x \,  x^2  e^{- x}  \sigma_{np}^{(2)}(xT).
}
We define the appropriately energy- and angle-averaged $np$ dipole scattering cross section  to be $ \langle \sigma_{np}^{(2)}(T) \rangle = \frac12 \int_0^\infty  d x\,e^{-x} x^2 \sigma_{np}^{(2)}(x)$, following~\cite{Rrapaj:2015wgs}, to further streamline the notation.

To make this useful for calculating the absorptive width for the dark photon we include the correct mixing angles and the dark photon on-shell condition $v^2 = 1-m'^2/\omega^2$. Thus, our final result for the absorptive width of the dark photon via inverse bremsstrahlung is
\begin{empheq}[box=\fbox]{align}  \label{Gamma-inv-calc}
\Gamma_{{\rm ibr},T}' = \frac{32}{3\pi} \frac{\alpha_{\rm EM} (\epm)_T^2 n_nn_p}{\omega^3} \pL\frac{\pi T}{m_N}\pR^{3/2}  \langle \sigma_{np}^{(2)} (T) \rangle, ~~~~ \Gamma_{{\rm ibr},L} '= \frac{ (\epm)_L^2}{ (\epm)_T^2} \frac{m'^2}{\omega^2} \Gamma_{{\rm ibr},T} ,
\end{empheq}
which again is the same for production or absorption (although when considering total power produced we multiply by $S_{L,T}$ to account for the spin degeneracy of the transverse mode). Assuming that we can compare to~\cite{Rrapaj:2015wgs} by $\Gamma_{\rm mfp}^{\text{(\cite{Rrapaj:2015wgs})}} \leftrightarrow \left. \bL 2\Gamma_{{\rm ibr},T}'^{\rm(\Eq{Gamma-inv-calc})} +\Gamma_{{\rm ibr},L}'^{\rm(\Eq{Gamma-inv-calc})} \bR \right/3 $, we find that the result there for inverse bremsstrahlung is larger by a factor $\frac{9\pi}{4\sqrt{1-m'^2/\omega^2}} \sim \cO(10)$, depending on $m'$ and $\omega$.

For the production rate, the exact same calculation goes through as in the absorption rate because of the soft-radiation approximation. However, we stipulate that the dark-photon energy must not exceed the energy of the collision. This gives $\Gamma_{\rm br} = \Gamma_{\rm SRA} \Theta(T_{\rm CM} - \omega)$, where the step function must consistently be included before doing the integral over\footnote{We may alternately enforce this step function by only considering $\omega$ between $m'$ and $T_{\rm CM}$ in the integral over $\omega$ for the luminosity, as done in~\cite{Rrapaj:2015wgs}.} $T_{\rm CM}$. This takes $\int_0^\infty \to \int_{\omega/T}^\infty$ in the definition of $\langle \sigma_{np}^{(2)} \rangle$, so we define $ \langle \sigma_{np}^{(2)} (\omega,T) \rangle = \frac12 \int_{\omega/T}^\infty dx x^2 e^{-x} \sigma_{np}^{(2)}(xT)$ and subsequently
\begin{empheq}[box=\fbox]{align}  \label{Gamma-prod-calc}
\Gamma_{{\rm br},T}' = \frac{32}{3\pi} \frac{\alpha_{\rm EM} (\epm)_T^2 n_nn_p}{\omega^3} \pL\frac{\pi T}{m_N}\pR^{3/2}  \langle \sigma_{np}^{(2)} (\omega, T) \rangle, ~~~~ \Gamma_{{\rm br},L} '= \frac{ (\epm)_L^2}{ (\epm)_T^2} \frac{m'^2}{\omega^2} \Gamma_{{\rm br},T} .
\end{empheq}
Empirically, we find that this correction leads to $\sim \cO(1)$ higher luminosities (and thus $\sim 50\%$ stronger bounds) compared to taking a simple rescaling $\Gamma_{{\rm br}|L,T}' = e^{-\omega/T} \Gamma_{{\rm ibr}|L,T}'$, as would be expected in an exact calculation where we should recover detailed balance.

\subsection{Decay}
For decay to an electron positron pair, we start with the relation
\beq
\Gamma_{e^+e^-;L,T}' = \frac1{2\omega} \int\frac{d^3\vec p_1}{2E_1} \frac{1-f_{e^-}(E_1)}{(2\pi)^3} \int\frac{d^3\vec p_2}{2E_2} \frac{1-f_{e^+}(E_2)}{(2\pi)^3} (2\pi)^4 \delta^4\pL K^\mu -P_1^\mu -P_2^\mu \pR \oL{\mL \cM \mR}^2_{L,T}.
\eeq
Assuming that the positrons are not Pauli blocked and that the electron distribution function is Fermi-Dirac, this simplifies immediately to
\beq 
\Gamma_{e^+e^-;L,T}' = \frac1{32\pi^2\omega} \int\frac{d^3\vec p_1}{E_1E_2} \frac{\delta\pL \omega -E_1 -E_2\pR \oL{\mL \cM \mR}^2_{L,T}}{\exp\pL\frac{\mu_e-E_1}T\pR+1}.
\eeq
Using the projection operators from \Eq{projection-operators} and $S'_L=1,S'_T=2$, we have
\beq\nonumber
\oL{\mL \cM \mR}^2_{L,T}  =16 \pi \alpha_{\rm EM} (\epm)_{L,T}^2 \times \cbL \begin{array}{ll} m_e^2+E_1E_2 - \frac{(E_1 \omega - m'^2/2)(E_2 \omega - m'^2/2)}{\omega^2 - m'^2} & ~{\rm(transverse)} \\
m'^2 -2 E_1E_2 +\frac{2(E_1 \omega - m'^2/2)(E_2 \omega - m'^2/2)}{\omega^2 - m'^2} & ~{\rm(longitudinal)} \end{array} \right.,
\eeq
where $E_2=\omega-E_1$ is understood everywhere. We verify as expected that $\mL \cM \mR^2_T+\mL \cM \mR^2_L\propto m'^2+2m_e^2$ (with no independent spin average for the different polarizations in this sum). For ease of notation, we also define ${\sf m}(E_1)^2_{L,T} \equiv \oL{\mL \cM \mR}^2_{L,T} /16\pi \alpha_{\rm EM} m'^2 \ep_{L,T}^2$. Defining the common factor $z(E_1) = \frac{E_1E_2}{m'^2} - \pL \frac{E_1 \omega}{m'^2} - \frac12\pR \pL \frac{E_2 \omega}{m'^2} - \frac12 \pR \frac1{\omega^2 /m'^2 -1 } $, the dimensionless matrix elements squared simplify to
\beq \label{ee-matrix-element}
{\sf m}(E_1)^2_T = m_e^2/m'^2 + z(E_1),~~ ~~~~~{\sf m}(E_1)^2_L = 1 - 2z(E_1) ,
\eeq
which should be $\sim \cO(1)$ for most energies.

Respecting the cylindrical symmetry of the process, ${\sf m}^2_{L,T}$ have no dependence on the azimuthal angle of the outgoing particles. Doing the $d\phi$ integral, our expression becomes
\beq \label{mfp-intermediate}
\Gamma_{e^+e^-;L,T}' = \frac{\alpha_{\rm EM} (\epm)_{L,T}^2 m'^2}\omega \int\frac{|\vec p_1|^2 d |\vec p_1| d \cos \theta}{E_1E_2 \pL \frac{|\vec p_1|}{E_1}+\frac{|\vec p_1|-k \cos \theta}{E_2} \pR} \frac{\delta\pL |\vec p_1| - p_* \pR {\sf m}^2_{L,T}(E_1)}{\exp\pL\frac{\mu_e-E_1}T\pR+1},
\eeq
where the extra factor under the measure is from changing variables in the delta function, and $p_*$ is the momentum that satisfies the delta function. Relativistic kinematics gives
\beq \label{pstar}
p_*^2 = \frac14 m'^2\bL \bar v^2 \sin^2 \theta+ \pL \beta \gamma + \gamma \bar v \cos \theta\pR^2 \bR \implies p_- \leq p_* \leq p_+, \qquad p_\pm = \frac12 \omega \mL \beta \pm \bar v \mR,
\eeq
where $\bar v = \sqrt{1-4m_e^2/m'^2}$ and $\beta = \sqrt{1-m'^2/\omega^2}=|\vec k|/\omega$. We also observe that
\beq
P_2^\mu = K^\mu - P_1^\mu \implies \cos\theta = \frac{E_1 \omega - \frac12 m'^2}{|\vec k|p_*} \implies d \cos \theta = \frac{\frac12 E_1 m'^2 - m_e^2 \omega}{|\vec k| (E_1^2-m_e^2)^{3/2}} dE_1,
\eeq
where we have used $p_* = \sqrt{E_1^2-m_e^2}$. We plug back into \Eq{mfp-intermediate} and after some cancellations the result is
\begin{empheq}[box=\fbox]{align} \label{mfp-decay}
\Gamma_{e^+e^-;L,T}' = \frac{\alpha_{\rm EM} (\epm)_{L,T}^2 m'^2}{\sqrt{\omega^2-m'^2}} \int_{x_-}^{x_+} \frac{ {\sf m}(\omega x)^2_{L,T} \, d x}{\exp\pL\frac{\mu_e-\omega x}{T}\pR+1},
\end{empheq}
where the limits of integration are
\beq
x_\pm = \frac{\sqrt{m_e^2 +p_\pm^2}}\omega = \frac12 \pm \frac12 \sqrt{\pL1-\frac{4m_e^2}{m'^2}\pR\pL1-\frac{m'^2}{\omega^2}\pR }.
\eeq
\Eq{mfp-decay} is not singular at $\omega \to m'$ because in this limit the bounds of the integration approach one another to cut off the division by zero. We are not able to integrate \Eq{mfp-decay} in generality due to the complicated form of the matrix element and the Pauli factor.

\section{Details for Calculating the Optical Depth}
\label{opt-depth-details}

To calculate the optical depth for an outgoing dark photon, we must integrate its absorptive width along its path out of the proto-neutron star all the way to $R_{\rm far}$. In this appendix we describe some subtleties in this calculation.

In general, the integral for the optical depth requires knowledge of the angular distribution of stellar material. This integral will furthermore be sensitive not only to the radius of production but also to the angle each dark photon makes with respect to the radial direction as it leaves the star. For simplicity, we will assume spherical symmetry of stellar material, and we will correct for the distribution of possible path length as follows. If the path of a dark photon produced at radius $r$ is directed radially outward, the optical depth is simply
\beq
\tau_{\rm radial, \, out}(r) = \int_r^{R_{\rm far}} \Gamma_{\rm abs}(\tilde r) d \tilde r.
\eeq
If a dark photon produced at radius $r$ is directed radially inward towards the core, it traverses the distance to the center of the star once going inward and once going out. In this case, the optical depth is
\beq
\tau_{\rm radial, \, in}(r) =  \tau_{\rm radial \, out}(r)  + 2 \int_0^r \Gamma_{\rm abs}(\tilde r) d \tilde r.
\eeq
Dark photons that point inwards but do not cross through the core will not experience as large of a suppression. Thus averaging over the angles with which a dark photon may be emitted  with respect to the radial direction, we have heuristically
\beq
\langle \tau(r) \rangle =  \tau_{\rm radial \, out}(r)  + a(r) \int_0^r \Gamma_{\rm abs}(\tilde r) d \tilde r,
\eeq
where the correction factor $a(r)$ accounts for the densities and temperatures along the chord that is traversed and takes into account the angles which encounter the core. We expect that $a$ depends on $r$ because, {\it e.g.}, the fraction of solid angles that cross the core region (where the density is relatively high) will vary depending on how close to the core the dark photon is produced: for $r<R_c$ all paths go through the core, near $r \gtrsim R_c$ about half of the paths return through the core, but well outside $R_c$ this fraction is $\sim R_c/\sqrt{R_c^2+r^2}$. Given these estimates, we expect $a$ to rise monotonically from 0 at $r=0$ to something close to $R_c/R_\nu$ at $r=R_\nu$. In our numerical work, we will use the approximation
\beq \label{tau-geometric}
\tau(r) = \bL 1 - \frac{r(r-R_c)}{2R_\nu^2} \bR \tau_{\rm radial \, out}(r).
\eeq
A similar geometric factor multiplying $\tau_{\rm radial \, out}(0)$ is found to equal $3/\pi^2$ for an object of uniform density in \cite{1986rpa..book.....R}, and determined to be close to $0.3$ in the numerical work of \cite{Burrows:1990pk} in the leakage approximation. The prefactor in \Eq{tau-geometric} does not deviate far from one in practice.

We also comment on the upper bound of the integration for the optical depth. Because we are interested in the luminosity that reaches regions outside of the neutrinosphere (possibly as far as the shock radius), we must integrate the partial widths at radii beyond the profiles of the simulations to get the total optical depth. To account for this subtlety, we approximate the optical depth as
\alg{
\tau(r) &=\int_r^{R_{\rm far}} \Gamma_{\rm tot}(\tilde r) d \tilde r 
\\ & = \pL R_{\rm far} - R_\nu \pR \bL \Gamma_{e^+e^-}(R_\nu) +\Gamma_{\rm sC}(R_\nu) + \frac1{10} \Gamma_{\rm ibr} (R_\nu) \bR + \int_r^{R_\nu} \Gamma_{\rm tot}(\tilde r) .
}
The factor of $1/10$ in front of $\Gamma_{\rm ibr}$ is present to account for the sharp dependence on $r$ carried by $\Gamma_{\rm ibr}$. If $\Gamma_{\rm ibr} \propto r^{-a}$ for $a \gg1$, we would expect the contribution at $R_{\rm far}$ to be negligible compared to the region near $R_\nu$. However, the rate is unlikely to fall in this simple way out to arbitrarily high radii, so we adopt this approximation (which has a soft but nonzero scaling with $R_{\rm far}$) as a very rough estimate.
Fortunately, $\Gamma_{\rm ibr}$ does not dominate the optical depth 
anywhere near the upper bounds of the $\ep-m'$ parameter space, so our results are insensitive to this prescription.

\section{Comparison with Prior Work}
\label{app:compare}

As mentioned above, our constrained region goes to somewhat smaller values of $\ep$ compared to other results on the dark-photon model~\cite{Bjorken:2009mm,Dent:2012mx,Kazanas:2014mca,Rrapaj:2015wgs}. This is due to a number of partially offsetting factors, which we discuss in detail here.

The biggest change in our methodology, which leads to a qualitative change in the asymptotic scaling at masses below $\sim 10\mev$, is the inclusion of finite-temperature effects on the Standard Model photon self-energy. Thus, our limits are constant in the product $\ep^2m'^2$ rather than $m'$-independent. This has been included in limits on dark photons from other stars~\cite{An:2013yfc}, but our results are the first to include plasma effects on the constraints from SN1987A. Another important change is that we do not assume that $A'$ particles thermalize at high mixing angle. Finally, our maximum constrained mass is somewhat lower in the fiducial model (analogous to the conditions assumed by previous work) because this mass is determined by where the upper and lower boundaries (which are relatively steep functions of $m'$) intersect.

\begin{figure}[t]
\begin{center}
\includegraphics[width=0.8\textwidth]{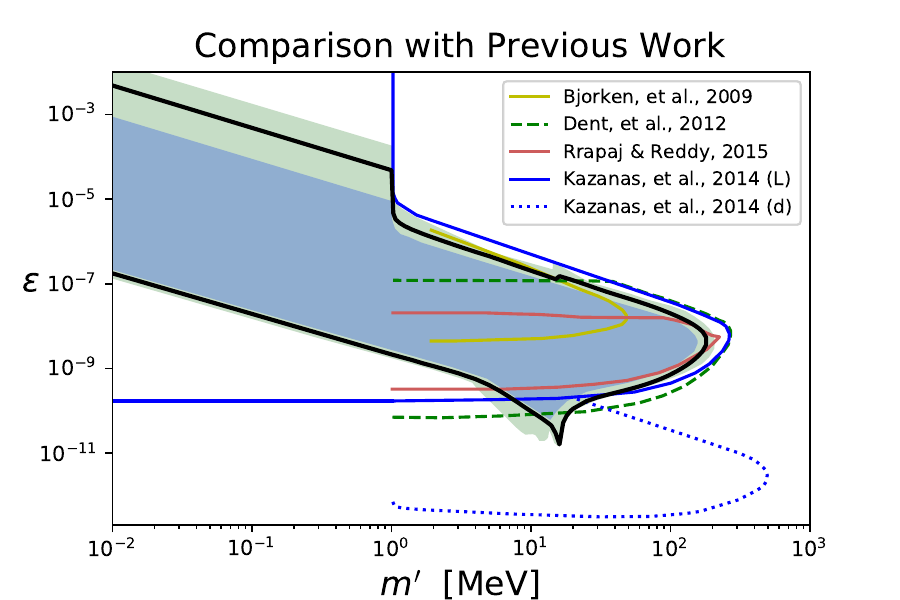}
\caption{Comparison of the bounds derived in this work with those from~\cite{Bjorken:2009mm,Dent:2012mx,Kazanas:2014mca,Rrapaj:2015wgs}. 
Note that the blue dotted line from~\cite{Kazanas:2014mca} is not based on anomalous cooling of the supernova (see text for details).}
\label{compare-SN-papers}
\end{center}
\end{figure}

The first (preliminary) bounds on this model space were presented in~\cite{Bjorken:2009mm}, using a simple rescaling of calculations in the axion case. This implicitly relies on a picture where a single pion exchange is used to model the nucleon interactions. The bounds in~\cite{Dent:2012mx,Kazanas:2014mca} also assume single pion exchange. As emphasized in~\cite{Rrapaj:2015wgs}, this is problematic across the entire range of energies considered in the proto-neutron star, and substantially overestimates the scattering cross section at the energy transfers expected in this environment. This leads to bounds that are too strong by almost an order of magnitude.

We also show a bound from~\cite{Kazanas:2014mca} for the range  $10^{-13} \lesssim \ep  \lesssim 10^{-10}$, denoted by a (d) on our plot. This bound relies on scaling arguments for the dark-photon production and assumptions about the outlying stellar material beyond the neutrinosphere. 
This estimate neglects finite-temperature effects, which we have shown are important in the core of the supernova below 10~MeV.  
We have not checked whether Pauli-blocking factors are important to include.  
Moreover, systematic uncertainties are at least as large as for the cooling bound.

\bibliography{SNbib.v2}

\providecommand{\href}[2]{#2}\begingroup\raggedright\begin{thebibliography}{10}

\bibitem{Jaeckel:2010ni}
J.~Jaeckel and A.~Ringwald, {\it {The Low-Energy Frontier of Particle
  Physics}},  {\em Ann. Rev. Nucl. Part. Sci.} {\bf 60} (2010) 405--437,
  [\href{http://www.arxiv.org/abs/1002.0329}{{\tt arXiv:1002.0329}}].

\bibitem{Hewett:2012ns}
J.~L. Hewett et~al., {\it {Fundamental Physics at the Intensity Frontier}},
  \href{http://www.arxiv.org/abs/1205.2671}{{\tt arXiv:1205.2671}}.

\bibitem{Essig:NLWCP:2013}
{Rouven Essig, John Jaros, William Wester (conveners) + many authors}, {\it
  {Summary Report of the New, Light Weakly-coupled Particles subgroup for
  Snowmass 2013}}, .

\bibitem{Alexander:2016aln}
J.~Alexander et~al., {\it {Dark Sectors 2016 Workshop: Community Report}},
  2016.
\newblock \href{http://www.arxiv.org/abs/1608.08632}{{\tt arXiv:1608.08632}}.

\bibitem{Raffelt:1996wa}
G.~G. Raffelt, {\em {Stars as laboratories for fundamental physics}}.
\newblock 1996.

\bibitem{Burrows:1986me}
A.~Burrows and J.~M. Lattimer, {\it {The birth of neutron stars}},  {\em
  Astrophys. J.} {\bf 307} (1986) 178--196.

\bibitem{Burrows:1987zz}
A.~Burrows and J.~M. Lattimer, {\it {Neutrinos from SN 1987A}},  {\em
  Astrophys. J.} {\bf 318} (1987) L63--L68.

\bibitem{future-DM-paper}
J.~H. Chang, R.~Essig, and S.~D. McDermott, {\it {Constraints on Light Hidden
  Sector Fermions from Supernova Cooling}}, .

\bibitem{Essig:2013vha}
R.~Essig, J.~Mardon, M.~Papucci, T.~Volansky, and Y.-M. Zhong, {\it
  {Constraining Light Dark Matter with Low-Energy $e^+e^-$ Colliders}},  {\em
  JHEP} {\bf 11} (2013) 167, [\href{http://www.arxiv.org/abs/1309.5084}{{\tt
  arXiv:1309.5084}}].

\bibitem{izaguirre:2013uxa}
E.~Izaguirre, G.~Krnjaic, P.~Schuster, and N.~Toro, {\it {New Electron
  Beam-Dump Experiments to Search for MeV to few-GeV Dark Matter}},  {\em Phys.
  Rev.} {\bf D88} (2013) 114015,
  [\href{http://www.arxiv.org/abs/1307.6554}{{\tt arXiv:1307.6554}}].

\bibitem{Dreiner:2013mua}
H.~K. Dreiner, J.-F. Fortin, C.~Hanhart, and L.~Ubaldi, {\it {Supernova
  constraints on MeV dark sectors from $e^+e^-$ annihilations}},  {\em Phys.
  Rev.} {\bf D89} (2014), no.~10 105015,
  [\href{http://www.arxiv.org/abs/1310.3826}{{\tt arXiv:1310.3826}}].

\bibitem{Bjorken:2009mm}
J.~D. Bjorken, R.~Essig, P.~Schuster, and N.~Toro, {\it {New Fixed-Target
  Experiments to Search for Dark Gauge Forces}},  {\em Phys. Rev.} {\bf D80}
  (2009) 075018, [\href{http://www.arxiv.org/abs/0906.0580}{{\tt
  arXiv:0906.0580}}].

\bibitem{Dent:2012mx}
J.~B. Dent, F.~Ferrer, and L.~M. Krauss, {\it {Constraints on Light Hidden
  Sector Gauge Bosons from Supernova Cooling}},
  \href{http://www.arxiv.org/abs/1201.2683}{{\tt arXiv:1201.2683}}.

\bibitem{Kazanas:2014mca}
D.~Kazanas, R.~N. Mohapatra, S.~Nussinov, V.~L. Teplitz, and Y.~Zhang, {\it
  {Supernova Bounds on the Dark Photon Using its Electromagnetic Decay}},  {\em
  Nucl. Phys.} {\bf B890} (2014) 17--29,
  [\href{http://www.arxiv.org/abs/1410.0221}{{\tt arXiv:1410.0221}}].

\bibitem{Rrapaj:2015wgs}
E.~Rrapaj and S.~Reddy, {\it {Nucleon-nucleon bremsstrahlung of dark gauge
  bosons and revised supernova constraints}},  {\em Phys. Rev.} {\bf C94}
  (2016), no.~4 045805, [\href{http://www.arxiv.org/abs/1511.09136}{{\tt
  arXiv:1511.09136}}].

\bibitem{Kapusta:2006pm}
J.~I. Kapusta and C.~Gale, {\em {Finite-temperature field theory: Principles
  and applications}}.
\newblock Cambridge University Press, 2011.

\bibitem{An:2013yfc}
H.~An, M.~Pospelov, and J.~Pradler, {\it {New stellar constraints on dark
  photons}},  {\em Phys. Lett.} {\bf B725} (2013) 190--195,
  [\href{http://www.arxiv.org/abs/1302.3884}{{\tt arXiv:1302.3884}}].

\bibitem{Hirata:1987hu}
{\bf Kamiokande-II} Collaboration, K.~Hirata et~al., {\it {Observation of a
  Neutrino Burst from the Supernova SN 1987a}},  {\em Phys. Rev. Lett.} {\bf
  58} (1987) 1490--1493. [,727(1987)].

\bibitem{Bionta:1987qt}
R.~M. Bionta et~al., {\it {Observation of a Neutrino Burst in Coincidence with
  Supernova SN 1987a in the Large Magellanic Cloud}},  {\em Phys. Rev. Lett.}
  {\bf 58} (1987) 1494.

\bibitem{Alekseev:1987ej}
E.~N. Alekseev, L.~N. Alekseeva, V.~I. Volchenko, and I.~V. Krivosheina, {\it
  {Possible Detection of a Neutrino Signal on 23 February 1987 at the Baksan
  Underground Scintillation Telescope of the Institute of Nuclear Research}},
  {\em JETP Lett.} {\bf 45} (1987) 589--592. [,739(1987)].

\bibitem{Turner:1987by}
M.~S. Turner, {\it {Axions from SN 1987a}},  {\em Phys. Rev. Lett.} {\bf 60}
  (1988) 1797.

\bibitem{Raffelt:1988py}
G.~G. Raffelt, {\it {SUPERNOVA SN1987A AND SOME PROPERTIES OF LIGHT, EXOTIC
  PARTICLES}},  in {\em {High-energy physics. Proceedings, 24th International
  Conference, Munich, Germany, August 4-10, 1988}}, 1988.

\bibitem{Kainulainen:1990bn}
K.~Kainulainen, J.~Maalampi, and J.~T. Peltoniemi, {\it {Inert neutrinos in
  supernovae}},  {\em Nucl. Phys.} {\bf B358} (1991) 435--446.

\bibitem{Kuflik:2012sw}
E.~Kuflik, S.~D. McDermott, and K.~M. Zurek, {\it {Neutrino Phenomenology in a
  3+1+1 Framework}},  {\em Phys. Rev.} {\bf D86} (2012) 033015,
  [\href{http://www.arxiv.org/abs/1205.1791}{{\tt arXiv:1205.1791}}].

\bibitem{Hanhart:2000er}
C.~Hanhart, D.~R. Phillips, S.~Reddy, and M.~J. Savage, {\it {Extra dimensions,
  SN1987a, and nucleon-nucleon scattering data}},  {\em Nucl. Phys.} {\bf B595}
  (2001) 335--359, [\href{http://www.arxiv.org/abs/nucl-th/0007016}{{\tt
  nucl-th/0007016}}].

\bibitem{Hanhart:2001fx}
C.~Hanhart, J.~A. Pons, D.~R. Phillips, and S.~Reddy, {\it {The Likelihood of
  GODs' existence: Improving the SN1987a constraint on the size of large
  compact dimensions}},  {\em Phys. Lett.} {\bf B509} (2001) 1--9,
  [\href{http://www.arxiv.org/abs/astro-ph/0102063}{{\tt astro-ph/0102063}}].

\bibitem{Dreiner:2003wh}
H.~K. Dreiner, C.~Hanhart, U.~Langenfeld, and D.~R. Phillips, {\it {Supernovae
  and light neutralinos: SN1987A bounds on supersymmetry revisited}},  {\em
  Phys. Rev.} {\bf D68} (2003) 055004,
  [\href{http://www.arxiv.org/abs/hep-ph/0304289}{{\tt hep-ph/0304289}}].

\bibitem{Bethe:1992fq}
H.~A. Bethe, {\it {Supernova 1987A: An Empirical and analytic approach}},  {\em
  Astrophys. J.} {\bf 412} (1993) 192--202.

\bibitem{Janka:2000bt}
H.~T. Janka, {\it {Conditions for shock revival by neutrino heating in core
  collapse supernovae}},  {\em Astron. Astrophys.} (2000)
  [\href{http://www.arxiv.org/abs/astro-ph/0008432}{{\tt astro-ph/0008432}}].
  [Astron. Astrophys.368,527(2001)].

\bibitem{Braaten:1993jw}
E.~Braaten and D.~Segel, {\it {Neutrino energy loss from the plasma process at
  all temperatures and densities}},  {\em Phys. Rev.} {\bf D48} (1993)
  1478--1491, [\href{http://www.arxiv.org/abs/hep-ph/9302213}{{\tt
  hep-ph/9302213}}].

\bibitem{Fischer:2016cyd}
T.~Fischer, S.~Chakraborty, M.~Giannotti, A.~Mirizzi, A.~Payez, and
  A.~Ringwald, {\it {Probing axions with the neutrino signal from the next
  galactic supernova}},  {\em Phys. Rev.} {\bf D94} (2016), no.~8 085012,
  [\href{http://www.arxiv.org/abs/1605.08780}{{\tt arXiv:1605.08780}}].

\bibitem{Nakazato}
K.~Nakazato.
\newblock Private communication.

\bibitem{Nakazato:2012qf}
K.~Nakazato, K.~Sumiyoshi, H.~Suzuki, T.~Totani, H.~Umeda, and S.~Yamada, {\it
  {Supernova Neutrino Light Curves and Spectra for Various Progenitor Stars:
  From Core Collapse to Proto-neutron Star Cooling}},  {\em Astrophys. J.
  Suppl.} {\bf 205} (2013) 2, [\href{http://www.arxiv.org/abs/1210.6841}{{\tt
  arXiv:1210.6841}}].

\bibitem{Hempel:2009mc}
M.~Hempel and J.~Schaffner-Bielich, {\it {Statistical Model for a Complete
  Supernova Equation of State}},  {\em Nucl. Phys.} {\bf A837} (2010) 210--254,
  [\href{http://www.arxiv.org/abs/0911.4073}{{\tt arXiv:0911.4073}}].

\bibitem{Liebendoerfer:2000fw}
M.~Liebendoerfer, A.~Mezzacappa, and F.-K. Thielemann, {\it {Conservative
  general relativistic radiation hydrodynamics in spherical symmetry and
  comoving coordinates}},  {\em Phys. Rev.} {\bf D63} (2001) 104003,
  [\href{http://www.arxiv.org/abs/astro-ph/0012201}{{\tt astro-ph/0012201}}].

\bibitem{Liebendoerfer:2001gu}
M.~Liebendoerfer, S.~Rosswog, and F.-K. Thielemann, {\it {An Adaptive grid,
  implicit code for spherically symmetric, general relativistic hydrodynamics
  in comoving coordinates}},  {\em Astrophys. J. Suppl.} {\bf 141} (2002)
  229--246, [\href{http://www.arxiv.org/abs/astro-ph/0106539}{{\tt
  astro-ph/0106539}}].

\bibitem{Liebendoerfer:2002xn}
M.~Liebendoerfer, O.~E.~B. Messer, A.~Mezzacappa, S.~W. Bruenn, C.~Y. Cardall,
  and F.~K. Thielemann, {\it {A Finite difference representation of neutrino
  radiation hydrodynamics for spherically symmetric general relativistic
  supernova simulations}},  {\em Astrophys. J. Suppl.} {\bf 150} (2004)
  263--316, [\href{http://www.arxiv.org/abs/astro-ph/0207036}{{\tt
  astro-ph/0207036}}].

\bibitem{Liebendoerfer:2003es}
M.~Liebendoerfer, M.~Rampp, H.~T. Janka, and A.~Mezzacappa, {\it {Supernova
  simulations with Boltzmann neutrino transport: A Comparison of methods}},
  {\em Astrophys. J.} {\bf 620} (2005) 840--860,
  [\href{http://www.arxiv.org/abs/astro-ph/0310662}{{\tt astro-ph/0310662}}].

\bibitem{Janka:2012wk}
H.-T. Janka, {\it {Explosion Mechanisms of Core-Collapse Supernovae}},  {\em
  Ann. Rev. Nucl. Part. Sci.} {\bf 62} (2012) 407--451,
  [\href{http://www.arxiv.org/abs/1206.2503}{{\tt arXiv:1206.2503}}].

\bibitem{Burrows:2012ew}
A.~Burrows, {\it {Colloquium: Perspectives on core-collapse supernova theory}},
   {\em Rev. Mod. Phys.} {\bf 85} (2013) 245,
  [\href{http://www.arxiv.org/abs/1210.4921}{{\tt arXiv:1210.4921}}].

\bibitem{Melson:2015spa}
T.~Melson, H.-T. Janka, R.~Bollig, F.~Hanke, A.~Marek, and B.~M{\"u}ller, {\it
  {Neutrino-driven Explosion of a 20 Solar-mass Star in Three Dimensions
  Enabled by Strange-quark Contributions to Neutrino--nucleon Scattering}},
  {\em Astrophys. J.} {\bf 808} (2015), no.~2 L42,
  [\href{http://www.arxiv.org/abs/1504.07631}{{\tt arXiv:1504.07631}}].

\bibitem{Sukhbold:2015wba}
T.~Sukhbold, T.~Ertl, S.~E. Woosley, J.~M. Brown, and H.~T. Janka, {\it
  {Core-Collapse Supernovae from 9 to 120 Solar Masses Based on
  Neutrino-powered Explosions}},  {\em Astrophys. J.} {\bf 821} (2016), no.~1
  38, [\href{http://www.arxiv.org/abs/1510.04643}{{\tt arXiv:1510.04643}}].

\bibitem{Blum:2016afe}
K.~Blum and D.~Kushnir, {\it {Neutrino Signal of Collapse-induced Thermonuclear
  Supernovae: the Case for Prompt Black Hole Formation in SN1987A}},  {\em
  Astrophys. J.} {\bf 828} (2016), no.~1 31,
  [\href{http://www.arxiv.org/abs/1601.03422}{{\tt arXiv:1601.03422}}].

\bibitem{Weldon:1983jn}
H.~A. Weldon, {\it {Simple Rules for Discontinuities in Finite Temperature
  Field Theory}},  {\em Phys. Rev.} {\bf D28} (1983) 2007.

\bibitem{Nyman:1968jro}
E.~M. Nyman, {\it {Soft-Photon Theory of Nucleon-Nucleon Bremsstrahlung}},
  {\em Phys. Rev.} {\bf 170} (1968), no.~5 1628.

\bibitem{Baier:1996vi}
R.~Baier, Y.~L. Dokshitzer, A.~H. Mueller, S.~Peigne, and D.~Schiff, {\it {The
  Landau-Pomeranchuk-Migdal effect in QED}},  {\em Nucl. Phys.} {\bf B478}
  (1996) 577--597, [\href{http://www.arxiv.org/abs/hep-ph/9604327}{{\tt
  hep-ph/9604327}}].

\bibitem{Pospelov:2008jk}
M.~Pospelov, A.~Ritz, and M.~B. Voloshin, {\it {Bosonic super-WIMPs as
  keV-scale dark matter}},  {\em Phys. Rev.} {\bf D78} (2008) 115012,
  [\href{http://www.arxiv.org/abs/0807.3279}{{\tt arXiv:0807.3279}}].

\bibitem{Redondo:2008ec}
J.~Redondo and M.~Postma, {\it {Massive hidden photons as lukewarm dark
  matter}},  {\em JCAP} {\bf 0902} (2009) 005,
  [\href{http://www.arxiv.org/abs/0811.0326}{{\tt arXiv:0811.0326}}].

\bibitem{Essig:2013goa}
R.~Essig, E.~Kuflik, S.~D. McDermott, T.~Volansky, and K.~M. Zurek, {\it
  {Constraining Light Dark Matter with Diffuse X-Ray and Gamma-Ray
  Observations}},  {\em JHEP} {\bf 11} (2013) 193,
  [\href{http://www.arxiv.org/abs/1309.4091}{{\tt arXiv:1309.4091}}].

\bibitem{Fradette:2014sza}
A.~Fradette, M.~Pospelov, J.~Pradler, and A.~Ritz, {\it {Cosmological
  Constraints on Very Dark Photons}},  {\em Phys. Rev.} {\bf D90} (2014), no.~3
  035022, [\href{http://www.arxiv.org/abs/1407.0993}{{\tt arXiv:1407.0993}}].

\bibitem{Berger:2016vxi}
J.~Berger, K.~Jedamzik, and D.~G.~E. Walker, {\it {Cosmological Constraints on
  Decoupled Dark Photons and Dark Higgs}},
  \href{http://www.arxiv.org/abs/1605.07195}{{\tt arXiv:1605.07195}}.

\bibitem{Bjorken:1988as}
J.~D. Bjorken, S.~Ecklund, W.~R. Nelson, A.~Abashian, C.~Church, B.~Lu, L.~W.
  Mo, T.~A. Nunamaker, and P.~Rassmann, {\it {Search for Neutral Metastable
  Penetrating Particles Produced in the SLAC Beam Dump}},  {\em Phys. Rev.}
  {\bf D38} (1988) 3375.

\bibitem{Riordan:1987aw}
E.~M. Riordan et~al., {\it {A Search for Short Lived Axions in an Electron Beam
  Dump Experiment}},  {\em Phys. Rev. Lett.} {\bf 59} (1987) 755.

\bibitem{Bross:1989mp}
A.~Bross, M.~Crisler, S.~H. Pordes, J.~Volk, S.~Errede, and J.~Wrbanek, {\it {A
  Search for Shortlived Particles Produced in an Electron Beam Dump}},  {\em
  Phys. Rev. Lett.} {\bf 67} (1991) 2942--2945.

\bibitem{Batell:2009yf}
B.~Batell, M.~Pospelov, and A.~Ritz, {\it {Probing a Secluded U(1) at
  B-factories}},  {\em Phys. Rev.} {\bf D79} (2009) 115008,
  [\href{http://www.arxiv.org/abs/0903.0363}{{\tt arXiv:0903.0363}}].

\bibitem{Blumlein:2011mv}
J.~Blumlein and J.~Brunner, {\it {New Exclusion Limits for Dark Gauge Forces
  from Beam-Dump Data}},  {\em Phys.Lett.} {\bf B701} (2011) 155--159,
  [\href{http://www.arxiv.org/abs/1104.2747}{{\tt arXiv:1104.2747}}].

\bibitem{Andreas:2012mt}
S.~Andreas, C.~Niebuhr, and A.~Ringwald, {\it {New Limits on Hidden Photons
  from Past Electron Beam Dumps}},  {\em Phys.Rev.} {\bf D86} (2012) 095019,
  [\href{http://www.arxiv.org/abs/1209.6083}{{\tt arXiv:1209.6083}}].

\bibitem{Pospelov:2008zw}
M.~Pospelov, {\it {Secluded U(1) below the weak scale}},  {\em Phys. Rev.} {\bf
  D80} (2009) 095002, [\href{http://www.arxiv.org/abs/0811.1030}{{\tt
  arXiv:0811.1030}}].

\bibitem{Reece:2009un}
M.~Reece and L.-T. Wang, {\it {Searching for the Light Dark Gauge Boson in
  Gev-Scale Experiments}},  {\em JHEP} {\bf 0907} (2009) 051,
  [\href{http://www.arxiv.org/abs/0904.1743}{{\tt arXiv:0904.1743}}].

\bibitem{Aubert:2009cp}
{\bf BABAR} Collaboration, B.~Aubert et~al., {\it {Search for Dimuon Decays of
  a Light Scalar Boson in Radiative Transitions $\Upsilon \to \gamma A_0$}},
  {\em Phys. Rev. Lett.} {\bf 103} (2009) 081803,
  [\href{http://www.arxiv.org/abs/0905.4539}{{\tt arXiv:0905.4539}}].

\bibitem{Archilli:2011zc}
{\bf KLOE-2} Collaboration, F.~Archilli et~al., {\it {Search for a vector gauge
  boson in $\phi$ meson decays with the KLOE detector}},  {\em Phys. Lett.}
  {\bf B706} (2012) 251--255, [\href{http://www.arxiv.org/abs/1110.0411}{{\tt
  arXiv:1110.0411}}].

\bibitem{Abrahamyan:2011gv}
{\bf APEX} Collaboration, S.~Abrahamyan et~al., {\it {Search for a New Gauge
  Boson in Electron-Nucleus Fixed-Target Scattering by the APEX Experiment}},
  {\em Phys. Rev. Lett.} {\bf 107} (2011) 191804,
  [\href{http://www.arxiv.org/abs/1108.2750}{{\tt arXiv:1108.2750}}].

\bibitem{Merkel:2011ze}
{\bf A1} Collaboration, H.~Merkel et~al., {\it {Search for Light Gauge Bosons
  of the Dark Sector at the Mainz Microtron}},  {\em Phys. Rev. Lett.} {\bf
  106} (2011) 251802, [\href{http://www.arxiv.org/abs/1101.4091}{{\tt
  arXiv:1101.4091}}].

\bibitem{Merkel:2014avp}
H.~Merkel et~al., {\it {Search at the Mainz Microtron for Light Massive Gauge
  Bosons Relevant for the Muon g-2 Anomaly}},  {\em Phys. Rev. Lett.} {\bf 112}
  (2014), no.~22 221802, [\href{http://www.arxiv.org/abs/1404.5502}{{\tt
  arXiv:1404.5502}}].

\bibitem{Babusci:2012cr}
{\bf KLOE-2} Collaboration, D.~Babusci et~al., {\it {Limit on the production of
  a light vector gauge boson in phi meson decays with the KLOE detector}},
  {\em Phys. Lett.} {\bf B720} (2013) 111--115,
  [\href{http://www.arxiv.org/abs/1210.3927}{{\tt arXiv:1210.3927}}].

\bibitem{Babusci:2014sta}
{\bf KLOE-2} Collaboration, D.~Babusci et~al., {\it {Search for light vector
  boson production in $e^+e^- \rightarrow \mu^+ \mu^- \gamma$ interactions with
  the KLOE experiment}},  {\em Phys. Lett.} {\bf B736} (2014) 459--464,
  [\href{http://www.arxiv.org/abs/1404.7772}{{\tt arXiv:1404.7772}}].

\bibitem{Lees:2014xha}
{\bf BaBar} Collaboration, J.~P. Lees et~al., {\it {Search for a Dark Photon in
  $e^+e^-$ Collisions at BaBar}},  {\em Phys. Rev. Lett.} {\bf 113} (2014),
  no.~20 201801, [\href{http://www.arxiv.org/abs/1406.2980}{{\tt
  arXiv:1406.2980}}].

\bibitem{Batley:2015lha}
{\bf NA48/2} Collaboration, J.~R. Batley et~al., {\it {Search for the dark
  photon in $\pi^0$ decays}},  {\em Phys. Lett.} {\bf B746} (2015) 178--185,
  [\href{http://www.arxiv.org/abs/1504.00607}{{\tt arXiv:1504.00607}}].

\bibitem{::2016lwm}
{\bf KLOE-2} Collaboration, A.~Anastasi et~al., {\it {Limit on the production
  of a new vector boson in $\mathrm{e^+ e^-}\rightarrow {\rm U}\gamma$,
  U$\rightarrow \pi^+\pi^-$ with the KLOE experiment}},  {\em Phys. Lett.} {\bf
  B757} (2016) 356--361, [\href{http://www.arxiv.org/abs/1603.06086}{{\tt
  arXiv:1603.06086}}].

\bibitem{Babusci:2015zda}
{\bf KLOE-2} Collaboration, A.~Anastasi et~al., {\it {Search for dark
  Higgsstrahlung in $e^{+}e^{?} \to \mu^{+}\mu^?$ and missing energy events
  with the KLOE experiment}},  {\em Phys. Lett.} {\bf B747} (2015) 365--372,
  [\href{http://www.arxiv.org/abs/1501.06795}{{\tt arXiv:1501.06795}}].

\bibitem{1986rpa..book.....R}
G.~B. {Rybicki} and A.~P. {Lightman}, {\em {Radiative Processes in
  Astrophysics}}.
\newblock June, 1986.

\bibitem{Burrows:1990pk}
A.~Burrows, M.~T. Ressell, and M.~S. Turner, {\it {Axions and SN1987A: Axion
  trapping}},  {\em Phys. Rev.} {\bf D42} (1990) 3297--3309.

\end{thebibliography}\endgroup
\bibliographystyle{JHEP}

\end{document}